\documentclass[final,3p,a4paper,sort&compress]{elsarticle}

\usepackage{amsmath,amssymb,amsthm}
\usepackage{graphicx}
\usepackage{textcomp}

\DeclareMathOperator{\sign}{sign}

\theoremstyle{definition}
\newtheorem{prop}{Proposition}

\begin{document}

\title{Equivalent continuous and discrete realizations of L\'evy flights:\\Model of one-dimensional motion of inertial particle}

\author{Ihor Lubashevsky}
\ead{i-lubash@u-aizu.ac.jp}
\address{University of Aizu, Ikki-machi, Aizu-Wakamatsu, Fukushima 965-8560, Japan}

\begin{abstract}
The paper is devoted to the relationship between the continuous Markovian description of L\'evy flights developed previously (\textit{Lubashevsky et al., Phys. Rev. E \textbf{79} (2009) 011110, \textbf{80} (2009) 031148;  Eur. Phys. J. B \textbf{78} (2010) 207, \textbf{82} (2011) 189}) and their equivalent representation in terms of discrete steps of a wandering particle, a certain generalization of continuous time random walks.
To simplify understanding the key points of the technique to be created, our consideration is confined to the one-dimensional model for continuous random motion of a particle with inertia. Its dynamics governed by stochastic self-acceleration is described as motion on the phase plane $\{x,v\}$ comprising the position $x$ and velocity $v=dx/dt$ of the given particle. A notion of random walks inside a certain neighborhood $\mathcal{L}$ of the line $v=0$ (the $x$-axis) and outside it is developed. It enables us to represent a continuous trajectory of particle motion on the plane $\{x,v\}$ as a collection of the corresponding discrete steps. Each of these steps matches one complete fragment of the velocity fluctuations originating and terminating at the ``boundary'' of $\mathcal{L}$. As demonstrated, the characteristic length of particle spatial displacement is mainly determined by velocity fluctuations with large amplitude, which endows the derived random walks along the $x$-axis with the characteristic properties of L\'evy flights. 
Using the developed classification of random trajectories a certain parameter-free core stochastic process is constructed. Its peculiarity is that all the characteristics of L\'evy flights similar to the exponent of the L\'evy scaling law are no more than the parameters of the corresponding transformation from the particle velocity $v$ to the related variable of the core process. In this way the previously found validity of the continuous Markovian model for all the regimes of L\'evy flights is explained.  Based on the obtained results an efficient ``single-peak'' approximation is constructed. In particular, it enables us to calculate the basic characteristics of L\'evy flights using the probabilistic properties of extreme velocity fluctuations and the shape of the most probable trajectory of particle motion within such extreme fluctuations. 
\end{abstract}


\maketitle

\section{Introduction}

During the last two decades there has been a great deal of research into L\'evy type stochastic processes in various systems (for a review see, e.g., Ref.~\cite{CTRW}). According to the accepted classification of the L\'evy type transport phenomena \cite{CTRW}, L\'evy flights are Markovian random walks characterized by the divergence of the second moment of walker displacement $x(t)$, i.e., $\left<[x(t)]^2\right> \to \infty$ for any time scale $t$. It is caused by a power-law asymptotics of the distribution function $\mathcal{P}(x,t)$. For example, in the one-dimensional case this distribution function exhibits the asymptotic behavior $\mathcal{P}(x,t)\sim [\overline{x}(t)]^\alpha/x^{1+\alpha}$ for $x\gg\overline{x}(t)$, where $\overline{x}(t)$ is the characteristic length of the walker displacements during the time interval $t$ and the exponent $\alpha$ belongs to the interval $0<\alpha<2$. The time dependence of the quantity $\overline{x}(t)$ obeys the scaling law $\overline{x}(t)\propto t^{1/\alpha}$. L\'evy flights are met, for instance, in the motion of tracer particles in turbulent flows \cite{Swinney}, the diffusion of particles in random media \cite{Bouchaud}, human travel behavior and spreading of epidemics \cite{Brockmann}, or economic time series in finance \cite{Stanley}.

As far as the developed techniques of modeling such stochastic processes are concerned, worthy of mention are, in particular, the Langevin equation with L\'evy noise (see, e.g., Ref.~\cite{Weron}) and the corresponding Fokker-Planck equations \cite{Schertzer1,Schertzer2,CiteNew1,CN100}, the description of anomalous diffusion with power-law distributions of spatial and temporal steps \cite{Fogedby1,Sokolov}, L\'evy flights in heterogeneous media \cite{Fogedby2,Honkonen,BrockmannGeisel,citeNNN3,citeNNN4} and in external fields \cite{BrockmannSokolov,Fogedby3},  constructing the Fokker-Planck equation for L\'evy type processes in nonhomogeneous media~\cite{CiteNew2,CiteNew3,CiteNew4}, first passage time analysis and escaping problem for L\'evy flights \cite{fptp1,fptp1Chech,fptp2,fptp3,fptp4,fptp5,fptp6,CiteNew5,CiteNew6,citeNNN1,citeNNN2}.

One of the widely used approaches to coping with L\'evy flights, especially in complex environment, is the so-called continuous time random walks (CTRW) \cite{CTRW1,CTRW2}. It models, in particular, a general class of L\'evy type stochastic processes described by the fractional Fokker-Planck equation \cite{CTRW3}. Its pivot point is the representation of a stochastic process at hand as a collection of random jumps (steps) $\{\delta\mathbf{x}, \delta t\}$  of a wandering particle in space and time as well. 
In the frameworks of the coupled CTRW the particle is assumed to move uniformly along a straight line connecting the initial and terminal points of one step. In this case the discrete collection of steps is converted into a continuous trajectory and the velocity $\mathbf{v}=\delta\mathbf{x}/\delta t$ of motion within one step is introduced. As a result, the given stochastic process is described by the probabilistic properties, e.g., of the collection of random variables $\{\mathbf{v},\delta t \}$. 
A more detailed description of particle motion lies beyond the CTRW model. Unfortunately, for L\'evy flights fine details of the particle motion within one step can be important especially in heterogeneous media or systems with boundaries because of the divergence of the moment  $\left<[\delta\mathbf{x}(\delta t)]^2\right>$. Broadly speaking, it is due to a L\'evy particle being able to jump over a long distance during a short time. The fact that L\'evy flights can exhibit nontrivial properties on scales of one step was demonstrated in Ref.~\cite{CiteNew5} studied the first passage time problem for L\'evy flights based on the leapover statistics.

Previously a new approach to tackling this problem was proposed  \cite{we1,we2,we3,we33}. It is based on the following nonlinear stochastic differential equation with white noise $\xi(t)$
\begin{equation}\label{int:1}
  \tau\frac{dv}{dt} = - \lambda v + \sqrt{\tau\big(v_a^2+v^2\big)} \xi(t)
\end{equation}
governing random motion of a particle wandering, e.g., in the one-dimensional space $\mathbb{R}_x$. Here $v=dx/dt$ is the particle velocity, the time scale $\tau$ characterizes the delay in variations of the particle velocity which is caused by the particle inertia, $\lambda$ is a dimensionless friction coefficient. The parameter $v_a$ quantifies the relative contribution of the additive component $\xi_a(t)$ of the Langevin force 
with respect to the multiplicative one $v \xi_m(t)$ which are combined within one term 
\begin{equation}\label{int:2}
   v_a\xi_a(t)+ v\xi_m(t)\Rightarrow\sqrt{\big(v_a^2+v^2\big)} \xi(t)\,.
\end{equation}
Here we have not specified the type of the stochastic differential equation \eqref{int:1} because in the given case  all the types are interrelated via the renormalization of the friction coefficient $\lambda$. It should be noted that models similar to Eq.~\eqref{int:1} within replacement \eqref{int:2} can be classified as the generalized Cauchy stochastic process \cite{Konno} and has been employed to study stochastic behavior of various nonequilibrium systems, in particular, lasers \cite{22}, on-off intermittency \cite{26}, economic activity \cite{27}, passive scalar field advected by fluid \cite{28}, etc. 

Model~\eqref{int:1} generates continuous Markovian trajectories obeying the L\'evy statistics on time scales $t\gg\tau$ \cite{we1,we2,we3}.  Using a special singular perturbation technique \cite{we2} it was rigorously proved for the superdiffusive regime matching $1<\alpha<2$ \cite{we1,we2} and also verified numerically for the quasiballistic ($\alpha = 1$) and superballistic ($0<\alpha<1$) regimes \cite{we3}. Moreover, the main expressions obtained for the distribution function $\mathcal{P}(x,t)$ and the scaling law $\overline{x}(t)$ within the interval $1< \alpha < 2$ turn out to hold also for the whole region $0<\alpha <2$  \cite{we3}.  After its generalization \cite{we33} model~\eqref{int:1} generates truncated L\'evy flights as well.

The given approach can be regarded as a continuous Markovian realization of L\'evy flights. Indeed, it is possible to choose the system parameters in such a way that, on one hand, the ``microscopic'' time scale $\tau$ be equal to  an arbitrary small value given beforehand and, on the other hand, the system behavior remain unchanged on scales $t\gg\tau$ \cite{we1,we2}. The goal of this paper is to elucidate the fundamental features of this approach, to explain the found validity of model~\eqref{int:1} for describing L\'evy flights of all the regimes, and to construct a certain generalization of the continuous time random walks that admits a rather fine representation of the particle motion within one step. The latter feature will demonstrate us a way to overcome the basic drawback of the CTRW model, it appeals to different mechanisms in modeling the particle motion within individual steps and the relative orientation of neighboring steps.

\section{Continuous Markov model of L\'evy flights}\label{1DModel} 

\subsection{Model}

Following \cite{we1,we3,we33} let us consider random walks $\{x(t)\}$ of an inertial particle wandering in the one-dimensional space $\mathbb{R}_x$ whose velocity $v=dx/dt$ is governed by the following stochastic differential equation written in the dimensionless form
\begin{equation}\label{sec1:eq1}
  \frac{dv}{dt}=-\alpha v k(v)+\sqrt{2}g(v)\circ\xi(t)\,.
\end{equation}
Here the constant $\alpha$ is a system parameter meeting the inequality
\begin{equation}\label{sec1:alpha}
	0<\alpha<2\,,
\end{equation}
the positive function $k(v)>0$ allows for nonlinear friction effects, $\xi(t)$ is the white Gaussian noise with the correlation function
\begin{equation}\label{sec1:wngf}
  \left\langle \xi(t)\xi(t')\right\rangle =\delta(t-t')\,,
\end{equation}
and whose intensity $g(v)>0$ depending on the magnitude the particle velocity describes the cumulative effect of the additive and multiplicative components of the Langevin random force as noted in Introduction. The coefficient $\sqrt2$ has been introduced for the sake of convenience. The units of the spatial and temporal scales have been chosen in such a manner that the equalities 
\begin{equation}\label{sec1:kg_v0}
	k(0) = 1\quad \text{and}\quad  g(0) = 1
\end{equation}
hold. Equation~\eqref{sec1:eq1} is written in the Stratonovich form, which is indicated with the multiplication symbol $\circ$ in the product of the white noise $\xi(t)$ and its intensity $g(v)$. To avoid possible misunderstanding we note that this equation has been written in the H\"anggi-Klimontovich form in Refs.~\cite{we1,we3,we33}. In the kinetic theory of gases quantities relative to $k(v)$ and $g(v)$ and describing regular and random  effects caused by the scattering of atoms or molecules are usually called ``kinetic'' coefficients (see, e.g., Ref.~\cite{LLKT}). For this reason the functions $k(v)$ and $g(v)$ together  
with other quantities to be derived from them 
will be also referred to as the kinetic coefficients.

In the present paper the main attention is focused on the special case when the kinetic coefficients $k(v)$ and $g(v)$ are of the form 
\begin{equation}\label{sec1:kg0}
	k_0(v) = 1\quad \text{and}\quad g_0(v) = \sqrt{1+v^2}
\end{equation}
and the random walks $\{x(t)\}$ generated by model~\eqref{sec1:eq1} can be classified as L\'evy flights for large time scales, i.e., for $t\gg1$ in the chosen units \cite{we1,we2,we3}.  However, where appropriate, the general form of these kinetic coefficients will be used to demonstrate a certain universality of the results to be obtained and the feasibility of their generalization, for example, to the truncated L\'evy flights \cite{we33}. Nevertheless, in further mathematical manipulations leading to particular results the following assumption about the behavior of the kinetic coefficients 
\begin{equation}\label{sec1:kgaa}
  k(v)\approx 1\,,\quad g(v)\approx v \quad\text{for}\quad 1\lesssim v \lesssim v_c\,, \quad\text{and}\quad 
  \frac{vk(v)}{g^2(v)}  > B  \quad\text{for}\quad v \gtrsim v_c
\end{equation}
will be accepted beforehand. Here $B>0$ is some positive constant and $v_c \gg 1$ is a certain critical velocity characterizing the region where the generated random walks deviate from L\'evy flights in properties. Naturally, case~\eqref{sec1:kg0} obeys these conditions in the limit $v_c\to\infty$.

To make it easier to compare the results to obtained in the following sections with the characteristic properties of the velocity fluctuations governed by Eq.~\eqref{sec1:eq1}, here let us find the stationary distribution function $P^\text{st}(v)$ of the particle velocity $v$. The Fokker-Planck equation describing the dynamics of the velocity distribution $P(v,t)$ and matching Eq.~\eqref{sec1:eq1} is written as (see, e.g., Ref.~\cite{MyBook})
\begin{equation}\label{sec1:FPv}
  \frac{\partial P}{\partial t} = \frac{\partial}{\partial v}\left\{ g(v) \frac{\partial[g(v) P]}{\partial v} + \alpha vk(v)P \right\}.
\end{equation}
Its stationary solution  $P^\text{st}(v)$ meets the equality
\begin{equation*}
        g(v) \frac{\partial[g(v) P^\text{st}]}{\partial v} + \alpha vk(v)P^\text{st} = 0 
\end{equation*}
whence it follows that
\begin{align}
\label{sec1:Pst}
	P^\text{st}(v) & = \frac{C_v}{g(v)}\exp\left[-\alpha \int\limits_0^v \frac{uk(u)}{g^2(u)}du\right]\,,\\
\intertext{where the constant}
\label{sec1:PstC}
    C_v &= \left\{2 \int\limits_{0}^{\infty}\frac{dv}{g(v)}\exp\left[-\alpha \int\limits_0^v \frac{uk(u)}{g^2(u)}du\right]\right\}^{-1}
\end{align}
is specified by the normalization of the distribution function $P(v,t)$ to unity.

\subsection{Additive noise representation}

To elucidate the general mechanism responsible for anomalous properties of the stochastic process at hand let us pass from the velocity $v$ to a new variable $\eta = \eta(v)$ introduced via the equation
\begin{equation}\label{sec2:u}
      \frac{d\eta}{dv} = \frac{1}{g(v)}\quad\text{subject to the condition}\quad \eta(0)= 0 \,.
\end{equation} 
Equation~\eqref{sec1:eq1} is of the Stratonovich form, thus, we may use the standard rules of change of variables in operating with it. Therefore relation~\eqref{sec2:u} between the variables $v$ and $\eta$  enables us to reduce Eq.~\eqref{sec1:eq1} to the following one
\begin{equation}\label{sec1:eq2}
  \frac{d\eta}{dt}=-\alpha\phi(\eta)+\sqrt{2}\xi(t)\,,
\end{equation}
%
where the function $\phi(\eta)$ is specified by the expression
\begin{equation}\label{sec1:phi}
  \phi(\eta) = \left. \frac{vk(v)}{g(v)}\right|_{v=v(\eta)}\,.
\end{equation}
The function $\phi(\eta)$ together with the parameter $\alpha$ determine the rate of the particle regular drift in the $\eta$-space, $\mathbb{R}_\eta$. In particular, in case~\eqref{sec1:kg0} we have 
\begin{equation}\label{sec1:voneta0}
   v = \sinh (\eta)\qquad\text{and}\qquad \phi_0(\eta) = \tanh(\eta)\,.
\end{equation}
It should be noted that the multiplication symbol $\circ$ has been omitted in Eq.~\eqref{sec1:eq2} because the noise $\xi(t)$ enters it additively and, thus, all the types of this stochastic differential equation have the same form (for details see, e.g., Ref.~\cite{MyBook}). Let us also introduce into consideration the potential 
\begin{equation}\label{sec1:Phi}
   \Phi(\eta) = \int\limits_0^\eta\phi(\zeta)d\zeta \equiv  \int\limits_0^{v(\eta)} \frac{u k(u)}{g^2(u)}du
\end{equation}
which will be used below; in case~\eqref{sec1:kg0} it can be written as 
\begin{equation}\label{sec1:Phi0}
   \Phi_0(\eta) = \ln\left[\cosh(\eta)\right]\,.
\end{equation}
Equation~\eqref{sec1:eq2} together with expression~\eqref{sec1:phi} enable us to find the time pattern $\{\eta(t)\}$ for a given realization of the noise $\{\xi(t)\}_{-\infty}^{+\infty}$.

Leaving expression~\eqref{sec2:u} aside for a moment and keeping in mind solely the governing equations~\eqref{sec1:eq1} and \eqref{sec1:eq2} we may state that the constructed stochastic process $\{\eta(t)\}$ and the random variations $\{v(t)\}$ of the particle velocity are related to each other via the white noise $\xi(t)$ and are of the same level of generality. Only relation~\eqref{sec2:u} treated as the definition of the function $\eta = \eta(v)$ causes us to regard the time variations $\{\eta(t)\}$ as a stochastic process derived from $\{v(t)\}$. However, expression~\eqref{sec2:u} may be read in the ``opposite way'' as the definition of the function $v = v(\eta)$. 
In this case the process $\{\eta(t)\}$ plays the role of the noise source endowing the particle motion with stochastic properties and the particle velocity $v=v(\eta)$ becomes a derivative random variable.  The particle displacement $x(t)$ in the space $\mathbb{R}_x$ during the time interval $t$  
\begin{equation}\label{sec1:Phi0x}
   x(t) = \int\limits_0^t v\left[\eta(t')\right]dt'
\end{equation}
is also a derivative stochastic process of $\{\eta(t)\}$. Integral~\eqref{sec1:Phi0x} can be treated in the Riemann sense for a given pattern $\{\eta(t)\}$ because the correlation function $\left<\eta(t)\eta(t')\right>$ of the stochastic process $\eta(t)$ is smooth, in particular, has no singularity at $t=t'$. It should be noted that the situation would be much more complex if we dealt with nonlinear integrals of white noise because the correlation function of white noise proportional to $\delta(t-t')$ comes up at the boundary of integration regions \cite{MyBook}.

As in the previous section let us find the stationary distribution function $p^\text{st}(\eta)$ of the variable $\eta$ which will be used below. Going in the same, way we appeal to the Fokker-Planck equation 
\begin{equation}\label{FPeta}
\frac{\partial p}{\partial t} = \frac{\partial }{\partial \eta}\left[\frac{\partial p}{\partial \eta} + \alpha \frac{d\Phi(\eta)}{d\eta}p\right]\,,
\end{equation}
matching Eq.~\eqref{sec1:eq2} and governing the evolution of the distribution function $p(\eta,t)$. In the limit $t\to\infty$ its solution gives us the desired result  
\begin{equation}\label{petast}
p^\text{st}(\eta) = \left[2\int\limits^{\infty}_{0}e^{-\alpha\Phi(\eta')} \,d\eta'\right]^{-1} e^{-\alpha\Phi(\eta)}\,.
\end{equation}
It should be noted that Exps.~\eqref{sec1:Pst} and \eqref{petast}, as it must, coincide with each other within the cofactor $d\eta/dv$ (see Exp.~\eqref{sec2:u}) coming from the transformation of the elementary volume in the transition $\mathbb{R}_v\to \mathbb{R}_\eta$.

\subsection{Trajectory classification}\label{sec:TC}

\begin{figure}
\begin{center}
\includegraphics[width=0.65\columnwidth]{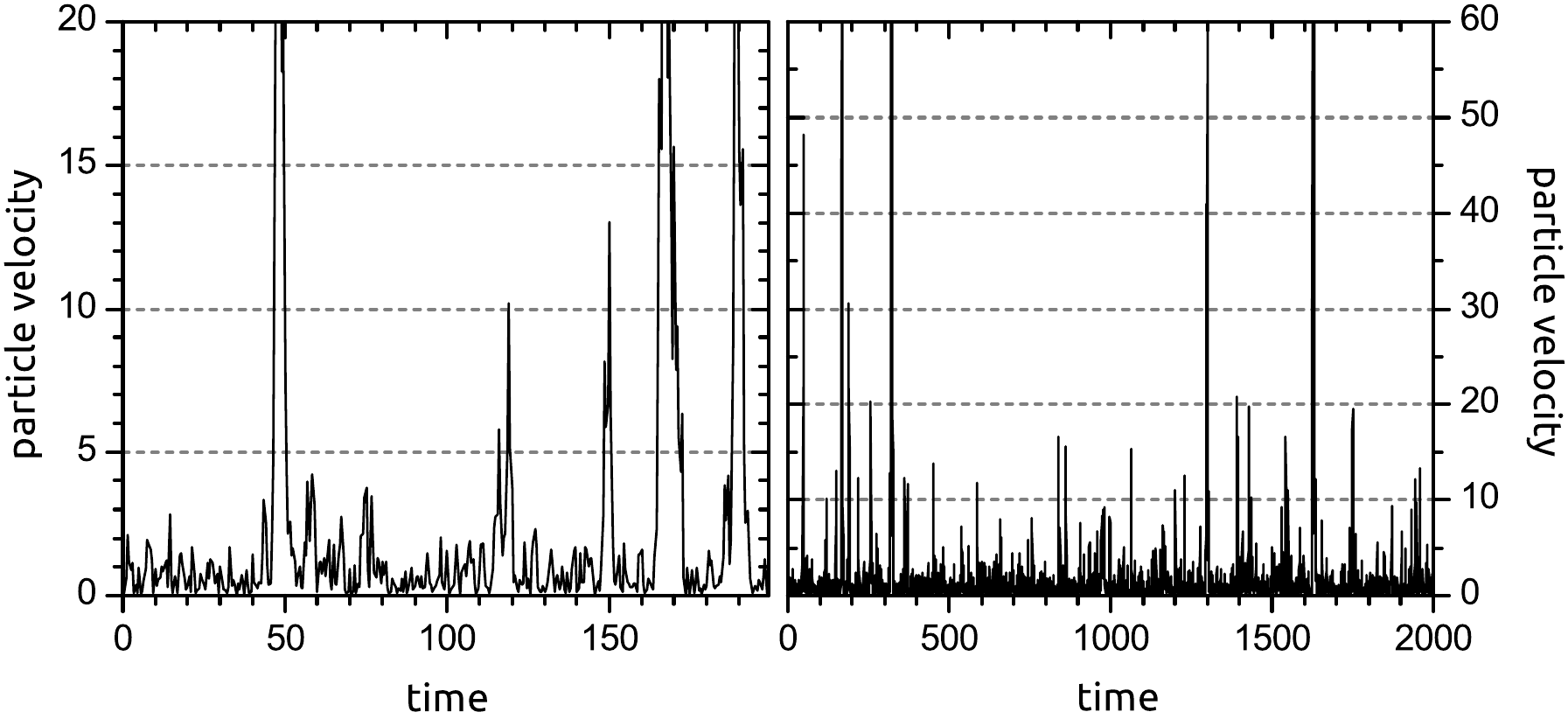}
\end{center}
\caption{The typical form of the time pattern $\{v(t)\}$ generated by model~\eqref{sec1:eq1} in the frameworks of ansatz~\eqref{sec1:kg0}. The two frames depict the same time pattern for different time scales. Based on the results presented in Ref.~\cite{we1}, the value $\alpha = 1.6$ was used.}
\label{F1}
\end{figure}

\begin{figure} 
\begin{center}
\includegraphics[width = 0.9\columnwidth]{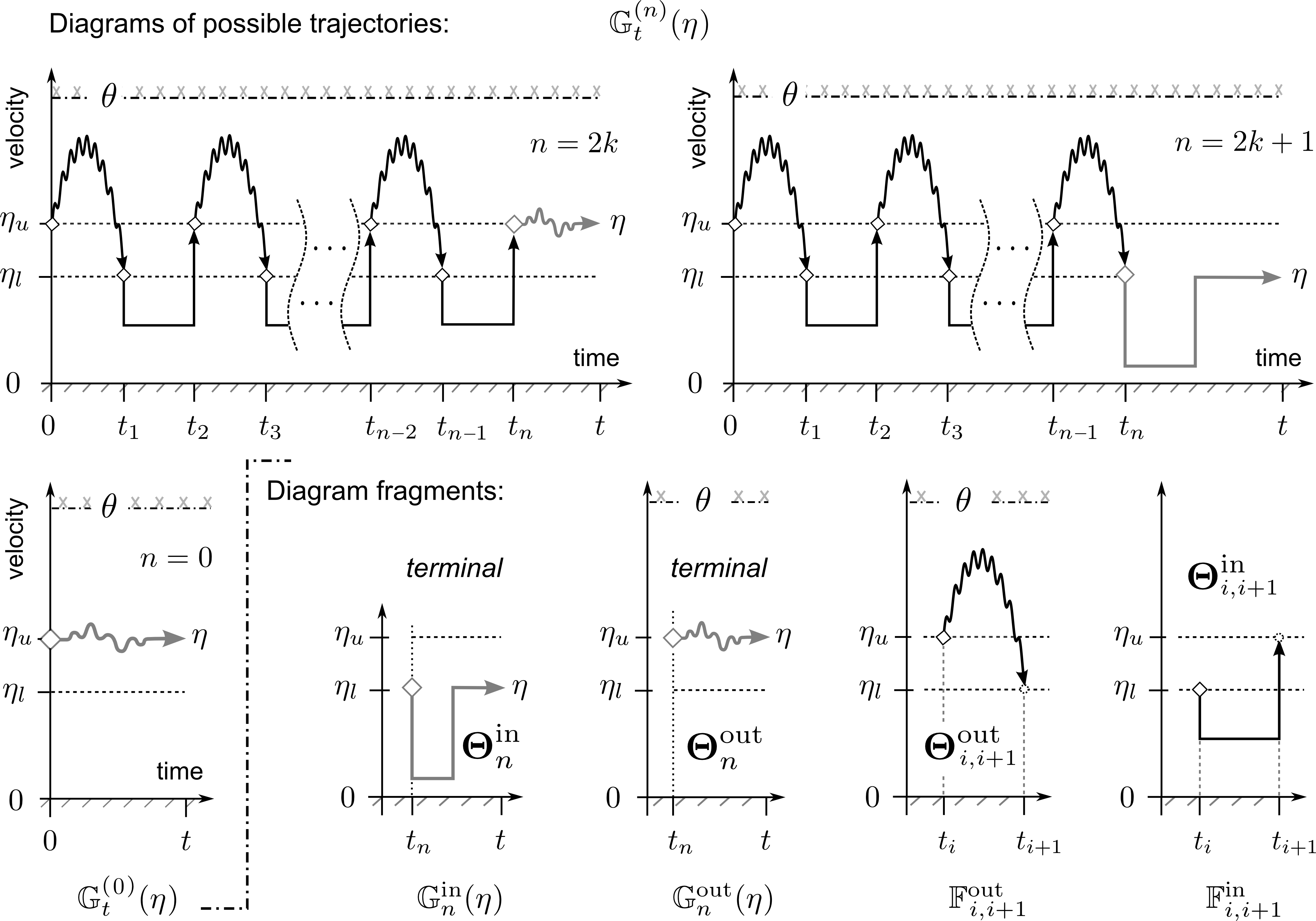}
\end{center}
\caption{The main diagrams and their basic fragments of the random walk classification enabling us to represent a particle trajectory 
in the space $\mathbb{R}_\eta$ as a countable sequence of alternate fragments matching the particle motion ``inside'' and ``outside'' a neighborhood $\mathcal{L}$ of the origin $\eta=0$. For the sake of simplicity the figure shows only trajectories starting from the  point $\eta_u$ and getting a certain point $\eta$ at time $t$.}
\label{F2}
\end{figure}

As shown in Refs.~\cite{we1,we2}, it is extreme fluctuations in the particle velocity, i.e., large amplitude peaks in the time pattern $\{v(t)\}$ that are responsible for the random walks at hand exhibiting the properties of L\'evy flights. 
Figure~\ref{F1} illustrates the multiscale structure of the pattern  $\{v(t)\}$. It turns out that for any time interval $t\gg1$ only a few peaks with the largest amplitudes inside it contribute mainly to the particle displacement \cite{we1}. The amplitudes of such velocity fluctuations are distributed according to the power-law and their characteristic magnitude grows with the observation time interval $t$ also according to the power-low. The former feature gives rise to the L\'evy distribution of the particle displacement, the latter one is responsible for the L\'evy time scaling \cite{we2}. 

These peculiarities make it attractive to employ the following classification of particle trajectories. The general idea is to partition any random trajectory $\{x(t)\}$ in such a manner that its fragments $\{x(t)\}_i^{i+1}$ or, more strictly, their ``projections'' $\{v(t)\}_i^{i+1}$ onto the velocity space $\mathbb{R}_v$ can be treated as random walks \textit{inside} or \textit{outside} a certain neighborhood $\mathcal{L}$ of the origin $v=0$. In this case it would be possible to regard the fragments of random walks outside the region $\mathcal{L}$ as individual peaks of the time pattern $\{v(t)\}$. Naturally, instead of the space $\mathbb{R}_v$ the space $\mathbb{R}_\eta$ of the variable $\eta$ can be equivalently used, so in what follows the corresponding neighborhood in the space $\mathbb{R}_\eta$ will be designated with the same symbol $\mathcal{L}$ to simplify notations. Besides, to operate with the desired partition efficiently it has to be countable. 

A direct implementation of this idea, however, faces a serious obstacle. The matter is that a random trajectory is not a smooth curve. Therefore, although for a wandering particle it is possible to calculate the probability of getting the boundary of $\mathcal{L}$ for the first time, the question about crossing this boundary for the second time is meaningless. The particle will cross the boundary immediately after the first one and ordering such intersection points in the countable manner seems to be just impossible.
The trajectory classification to be constructed below enables us to overcome this problem and, thus, to implement the desired partition. Previously the key elements of this classification were used in describing grain boundary diffusion \cite{Keigan1} and subdiffusion along a comb structure \cite{Keigan2}. 

Figure~\ref{F2} demonstrates the required classification of the particle random walks $\{\eta(t)\}$ in the diagram form.  Let us describe it step by step. 

(\textreferencemark)
First of all, just to  simplify subsequent mathematical manipulations we confine our consideration to the upper half-space $\mathbb{R}^+_\eta = \{\eta\geq 0\}$ assuming its boundary, $\eta = 0$, to be reflecting. The symmetry of the kinetic coefficient $\phi(\eta)$, namely, $\phi(-\eta) = \phi(\eta)$, allows it. Then two values $\eta_l$ and $\eta_u$ such that 
\begin{equation}\label{sec1:etaul}
 0 < \eta_l < \eta_u
\end{equation}
have to be introduced. The choice of their specific magnitudes will be reasoned below. Here, leaping ahead, we note that in the frameworks of ansatz~\eqref{sec1:kg0} the value $\eta_l$ should not be a too large number enabling us, nevertheless, to approximate the kinetic coefficient $\phi_0(\eta)=\tanh(\eta)$ by the sign-function for $|\eta|>\eta_l$,
\begin{equation}\label{sec1:sign}
    \phi_0(\eta)\approx 
    \sign(\eta) = 
    \begin{cases}
    1 &\text{if $\eta>0$}\,,\\
    0 &\text{if $\eta = 0$\,,}\\
    -1 &\text{if $\eta<0$}\,.
    \end{cases}
\end{equation}
For example, for $\eta_l = 2.0$ the estimates of $\tanh(\eta_l)\approx 0.96$ and $v_l = \sinh(\eta_l)\approx 3.6$ hold by virtue of \eqref{sec1:voneta0}. The value $\eta_u$ just should not exceed $\eta_l$ substantially. So in some sense
\begin{equation}\label{sec1:etaul0}
 1 \lesssim \eta_l \qquad\text{and}\qquad \eta_u -\eta_l \lesssim 1
\end{equation}
is the optimal choice. Certainly, the final results describing the particle displacement in the real space $\mathbb{R}_x$ do not depend on the two values. 

(\textreferencemark)
Again to simplify mathematical constructions without loss of generality let us consider trajectories $\{\eta(t')\}_{t'=0}^{t'=t}$ originating from the point $\eta_u$, i.e., set $\eta(t')|_{t'=0}=\eta_u$. Their terminal point $\eta = \eta(t')|_{t'=t}$ can take an arbitrary value. 

(\textreferencemark)
Now we are able to specify the desired partition. Each trajectory $\{\eta(t')\}_{t'=0}^{t'=t}$ is represented as a certain collection of fragments shown in Fig.~\ref{F2}, which is written symbolically as  
\begin{equation}\label{revis:1}
  \mathbb{G}_t^{(n)}(\eta) = \mathbb{F}_{01}^\text{out} \otimes \mathbb{F}_{12}^\text{in} \otimes \mathbb{F}_{23}^\text{out} \otimes \mathbb{F}_{34}^\text{in}\otimes\ldots\otimes \mathbb{G}_n^\text{out/in}(\eta)\,.
\end{equation}
If there is only one fragment in the collection $\mathbb{G}_t^{(n)}(\eta)$, namely, the last fragment, then the number $n$ of the internal fragments is set equal to zero, $n=0$. The meaning of these fragments is as follows.

\begin{itemize}
\item[$\circ$] The first fragment $\mathbb{F}^\text{out}_{01}$ represents the motion of the particle outside the interval $\mathcal{L}_l=[0,\eta_l)$ until it gets the point $\eta_l$ for the first time at a time moment $t_1>0$. Further such particle motion will be referred to as random walks \textit{outside} the neighborhood $\mathcal{L}$ of the origin $\eta=0$.

\item[$\circ$] The next fragment $\mathbb{F}^\text{in}_{12}$ matches the particle wandering inside the interval $\mathcal{L}_u=[0,\eta_u)$ until it gets the point $\eta_u$ for the first time at a certain moment $t_2>t_1$. Particle motion of this type will be referred to as random walks \textit{inside} the neighborhood $\mathcal{L}$.

\item[$\circ$] The following two fragments $\mathbb{F}^\text{out}_{23}$ and $\mathbb{F}^\text{in}_{34}$ are similar to the first and second ones, $\mathbb{F}^\text{out}_{01}$ and $\mathbb{F}^\text{in}_{12}$, respectively. The time moments corresponding to their terminal points are designated as $t_3$ and $t_4$. A sequence of such fragments alternating one another makes up the internal part of the given trajectory.

\item[$\circ$] Sequence~\eqref{revis:1} ends with a final fragment which can be of two configurations, $\mathbb{G}^\text{out}_{n}(\eta)$ and $\mathbb{G}^\text{in}_n(\eta)$. 
\begin{itemize}
\item[-] The former one, $\mathbb{G}^\text{out}_{n}(\eta)$, is related to the particle motion starting from the point $\eta_u$ at time $t_n$ and getting the point $\eta$ at time $t$ without touching the boundary point $\eta=\eta_l$ of the interval $\mathcal{L}_l$. This configuration exists only for $\eta > \eta_l$ and is characterized by an even number of the intermediate points, $n = 2k$, of the trajectory partition. 

\item[-] The latter one, $\mathbb{G}^\text{in}_n(\eta)$, is similar to the former configuration within the exchange of the start and boundary points; now $\eta_l$ is the start point whereas $\eta_u$ is the boundary point of the interval $\mathcal{L}_u$. For the configuration $\mathbb{G}^\text{in}_n(\eta)$ to exist the terminal point $\eta$ has to meet the inequality $\eta < \eta_u$. The corresponding number of the partition points is odd, $n = 2k+1$. 
\end{itemize}
It should be noted that when the terminal point $\eta$ of the given trajectory belongs to the interval $\eta_l < \eta<\eta_u$ both the configurations exist.

\item[$\circ$] In addition, for $\eta >\eta_l$ there is a special configuration $\mathbb{G}_t^\text{(0)}(\eta)$ of the trajectories that start from the point $\eta_u$ at the initial time $t=0$ and get the point $\eta$ at time $t$ without touching the boundary point of $\mathcal{L}_l$, i.e., $\eta = \eta_l$. It represents sequence~\eqref{revis:1} with no internal fragments, $n=0$. This configuration is actually equivalent to  $\mathbb{G}^\text{out}_n(\eta)$ with $t_n=0$. 

\end{itemize}

(\textreferencemark)
The introduced fragments of particle trajectories may be characterized by other classification parameters, denoted symbolically as $\boldsymbol{\Theta}^{\text{out/in}}_{i,i+1}$, in addition to the initial and terminal time moments of their realization, $t_i$ and $t_{i+1}$. Figure~\ref{F2} shows such parameters for the terminal fragments of $\mathbb{G}^{(n)}_t(\eta)$ too. The collection $\{\boldsymbol{\Theta}^{\text{out/in}}_{i,i+1}\}$ is determined by the specific details we want to know about the time pattern $\{\eta(t')\}_{t'=0}^{t'=t}$.  
In the present paper all the random walks in $\mathbb{R}_v$ will be separated into different groups considered individually according to the largest amplitude $\theta$ attained by the variable $\eta$ during the corresponding time interval $[0,t]$. As will be clear below, this classification can implemented imposing the addition requirement on each fragment of random walks outside the region $\mathcal{L}$ that bounds variations of the variable $\eta$ inside it. Namely, we assume the variable $\eta$ not to exceed $\theta$,
\begin{equation}\label{sec1:theta}
\eta_l < \eta(t) < \theta \qquad \text{for}\qquad t_i < t < t_{i+1}\quad\text{and}\quad\forall i\,.
\end{equation} 
In this case each set $\boldsymbol{\Theta}^\text{out}_{i,i+1} = \{\theta\}$ contains only one parameter taking the same value for all of them. For the terminal fragment $\mathbb{G}_n^\text{out}(\eta)$ a similar condition holds. For the random walks inside the region $\mathcal{L}$ there are no additional classification parameters, so the sets $\boldsymbol{\Theta}^\text{in}_{i,i+1}$ are empty. \hfill $\blacksquare$ 

The constructed partition enables us to develop a more sophisticated description of the pattern  $\{\eta(t)\}$, in particular, consider its configurations where the amplitudes of its $m$ largest peaks take given values $\theta_1,\theta_2,\ldots, \theta_m$.  For sure, such analysis will allow us to penetrate much deeper into the properties of L\'evy flights, which, however, is worthy of individual investigation.

The remaining part of the paper will be devoted to the statistical properties of random walks described in terms of the constructed partition. Before this, however, let us discuss the relationship between the given classification of the particle motion and the well-known model of continuous time random walks (CTRW). 

\subsection{Constructed partition as a generalized CTRW model}

The CTRW model imitates random motion of particles by assigning to each jump of a wandering particle a jump length $x$ and a waiting time $t$ elapsing between two successive jumps which are distributed with a certain joint probability density $\psi(x,t)$. When it is possible to write this probability density as the product of the individual probability densities of jump length, $\psi_x(x)$, and waiting time, $\psi_t(t)$, i.e.,  $\psi(x,t)= \psi_x(x)\psi_t(t)$, the two quantities can be regarded as independent random variables and the  model is called decoupled CTRW. 
Another widely met version of this model, coupled CTRW, writes the probability density $\psi(x,t)$ as the product of two functions $\psi(x,t)=\psi_x(x)\psi_v(x/t)$ or $\psi(x,t)=\psi_t(t)\psi_v(x/t)$. In the given case the jump parameters $x$ and $t$ are no longer independent random variables; the independent variables are the jump length $x$ (or the waiting time $t$) and the mean particle velocity $v= x/t$. The continuous implementation of the coupled CTRW is based on the assumption that within one jump the particle moves along the straight line connecting its initial and terminal points with the fixed velocity $v=x/t$.

The constructed partition can be regarded as a certain generalization of CTRW that allows one to consider the detailed structure of elementary steps. Indeed, a pair of succeeding fragments of random walks inside, $\mathbb{F}^\text{in}_{i,i+1}$, and outside, $\mathbb{F}^\text{out}_{i+1,i+2}$, the region $\mathcal{L}$ form an elementary step of the equivalent discrete random walks whose jump length and waiting time are random variables. 
If the random walks inside the region $\mathcal{L}$ do not contribute  substantially to the particle displacement $x$ we may speak about a stochastic process similar to coupled CTRW. The main difference between the two processes is due to the fact that the particle under consideration does not move uniformly within the fragment  $\mathbb{F}^\text{out}_{i+1,i+2}$. So the knowledge of the corresponding mean velocity is not enough to describe the real particle motion. The random walks inside the region $\mathcal{L}$ can be essential for the particle motion in space when, for example, a wandering particle spends the main time inside the region $\mathcal{L}$. In this case the corresponding stochastic process may be categorized as coupled-decoupled CTRW. In the case under consideration, as will be demonstrated below, the random walks inside the region $\mathcal{L}$ are not responsible for the L\'evy type behavior of the particle motion. However, the constructed partition actually does not require the normal behavior of random walks near the origin $v=0$, so it could be also used for modeling stochastic processes with kinetic coefficients exhibiting singularities at $v=0$. Under such conditions the random walks inside the region $\mathcal{L}$ are also able to cause anomalous properties of wandering particles, giving rise to the power-law distribution of the waiting time.

\subsection{Green function}

\begin{figure}
\begin{center}
\includegraphics[width = 0.4\columnwidth]{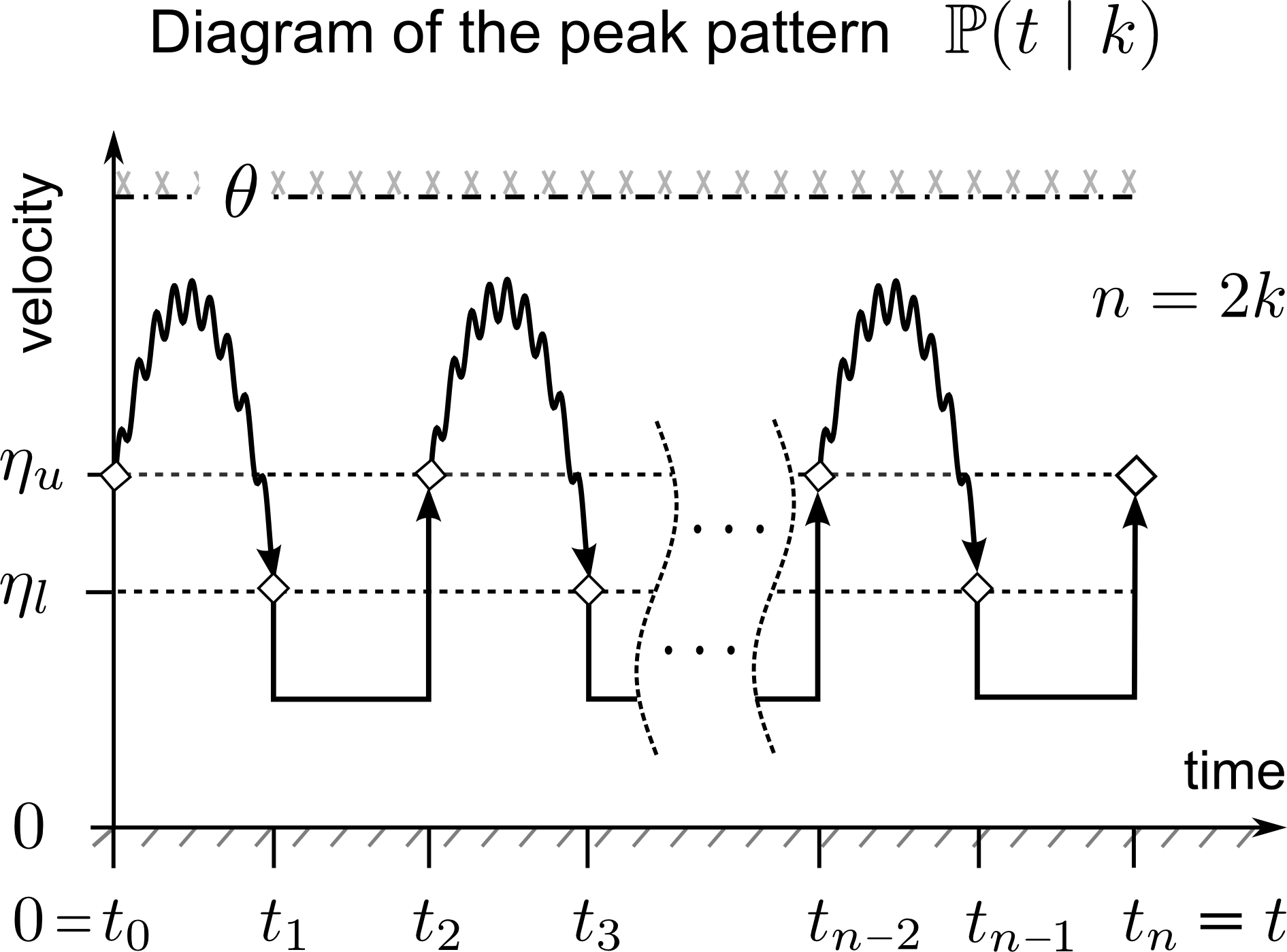} 
\end{center}
\caption{Diagram illustrating the structure of the peak pattern $\mathbb{P}(t\mid k)$ for $k >0$.}
\label{F3}
\end{figure}

Continuing the constructions of Sec.~\ref{sec:TC} the present section analyzes the statistical properties of the random trajectories $\{\eta(t')\}_{t'=0}^{t'=t}$ meeting condition~\eqref{sec1:theta} that are imposed upon all the fragments of the random walks outside the region $\mathcal{L}$. It will be done based on the calculation of the Green function $\mathcal{G}(\eta,t)$, i.e., the probability density of finding the particle at the point $\eta$ at time $t$ provided initially, $t=0$, it is located at the point $\eta_u$ and does not cross the boundary $\eta=\theta$ within the time interval $(0,t)$.  The developed classification of random walks illustrated by the diagrams in Fig.~\ref{F2} enables us to write 
\begin{equation}\label{sec2:G0}
    \mathcal{G}(\eta,t)=   \Theta_\text{H}(\eta -\eta_l) \mathcal{G}^\vartriangle(\eta,t) + 
                           \Theta_\text{H}(\eta_u -\eta) \mathcal{G}^\triangledown(\eta,t)\,,
\end{equation}
where we have introduced the functions
\begin{subequations}\label{sec2:Gab}
\begin{align}
\label{sec2:Ga}
    \mathcal{G}^\vartriangle(\eta,t) & = \sum_{k=0}^\infty \int\limits_0^t dt'  \mathcal{G}^\text{out}(\eta,t-t') \mathfrak{P}_k(t') \,,
\\
\label{sec2:Gb}
    \mathcal{G}^\triangledown(\eta,t) &=  \sum_{k=0}^\infty \int\limits_0^t dt' \int\limits_0^{t'} dt''
        \mathcal{G}^\text{in}(\eta,t-t') \mathcal{F}^\text{out}(t'-t'') \mathfrak{P}_k(t'') \,,
\end{align}
\end{subequations}
and the Heaviside step function  
\begin{equation}\label{sec2:Heav}
\Theta_\text{H}(\eta) = \begin{cases}
                  1, & \text{if $\eta > 0$},\\   0, & \text{if $\eta < 0$}
               \end{cases}
\end{equation}
to combine the cases $0\leq \eta < \eta_l$, $\eta_l < \eta < \eta_u$, and $\eta_u < \eta < \theta$ in one formula. Here $\mathcal{G}^\text{out}(\eta,t)$ is the probability density of finding the particle at the point $\eta$ at time $t$ provided it starts from the point $\eta_u$ remains inside the region $\eta_l < \eta<\theta$ within the time interval $0<t'<t$. The function $\mathcal{G}^\text{in}(\eta,t)$ is actually the same probability density within the replacement of the start point $\eta_u\to\eta_l$ and now the localization region is $0\leq \eta < \eta_u$.  
The function $\mathfrak{P}_k(t)$ specifies the probabilistic weight of the pattern $\mathbb{P}(t|k)$ shown in Fig.~\ref{F3} which comprises $k$ fragments of random walks outside the region $\mathcal{L}$ and $k$ fragments of random walks inside it. The former fragments will be referred also as to peaks of the pattern $\mathbb{P}(t|k)$. The probabilistic weight of the pattern  $\mathbb{P}(t|k)$ is given by the following formula for $k>0$
\begin{subequations}\label{sec2:G2}
\begin{align}
\label{sec2:G2a}
  \mathfrak{P}_k(t) &= \idotsint\limits_{0<t_1<t_2 \cdots < t_{2k-1} < t}  dt_1dt_2\ldots dt_{2k-1} 
                           \prod_{i=0}^{k-1} \mathcal{F}^\text{in}(t_{2i+2}-t_{2i+1}) \mathcal{F}^\text{out}(t_{2i+1}-t_{2i})
\\
\intertext{with $t_0=0$ and for $k = 0$, by definition,}
\label{sec2:G2b}
\mathfrak{P}_0(t) &= \delta(t)\,. 
\end{align}
\end{subequations}
In Exps.~\eqref{sec2:Gb} and \eqref{sec2:G2a} the functions $\mathcal{F}^\text{out}(t)$ and $\mathcal{F}^\text{in}(t)$ with the corresponding values of the argument $t$ determine the probabilistic weights of the fragments $\mathbb{F}^\text{out}_{i,i+1}$ and $\mathbb{F}^\text{in}_{i,i+1}$, respectively. The meaning of the former function is the probability density of reaching the point $\eta_l$ for the first time at the moment $t$ provided the particle is located initially at the point $\eta_u$. The meaning of the latter function is the same within the replacement $\eta_l \leftrightarrow \eta_u$. To simplify the notations, the quantities $\eta_u$, $\eta_l$, and $\theta$ have been omitted from the argument list of the corresponding functions noted above.

Integrals \eqref{sec2:Gab} and \eqref{sec2:G2a} are of the convolution type, therefore, it is convenient to pass from the time variable $t$ to the corresponding variable $s$ using the Laplace transform 
\begin{equation}\label{sec2:Lapl}
   W(s) := \mathcal{W}^L(s) =\int_0^\infty dt e^{-st}\mathcal{W}(t)\,,   
\end{equation}
where $\mathcal{W}(t)$ is one of the functions entering Exps.~\eqref{sec2:Gab} and \eqref{sec2:G2}. In these terms the given integrals are reduced to algebraic expressions, namely,
\begin{equation}\label{sec2:G2L}
  \mathfrak{P}^L_k(s) = \left[{F}^\text{in}(s) {F}^\text{out}(s)\right]^k\,.
\end{equation}
and, thus, 
\begin{subequations}\label{sec2:GabL}
\begin{align}
\label{sec2:GaL}
    {G}^\vartriangle(\eta,s) & = \sum_{k=0}^\infty \left[{F}^\text{in}(s) {F}^\text{out}(s)\right]^k {G}^\text{out}(\eta,s) 
                               = \frac{{G}^\text{out}(\eta,s)}{1-{F}^\text{in}(s) {F}^\text{out}(s)}\,,
\\
\label{sec2:GbL}
    {G}^\triangledown(\eta,s) &= \sum_{k=0}^\infty \left[{F}^\text{in}(s) {F}^\text{out}(s)\right]^k {F}^\text{out}(s){G}^\text{in}(\eta,s)
                               = \frac{{F}^\text{out}(s){G}^\text{in}(\eta,s)}{1-{F}^\text{in}(s) {F}^\text{out}(s)} 
\end{align}
\end{subequations}
because $|{F}^\text{in}(s) {F}^\text{out}(s)|<1$.

As demonstrated in \ref{App:inL} the functions $F^\text{in}(s),\, F^\text{out}(s)\approx 1$ for $s\ll 1$ provided, in addition, the inequality $\theta \gg 1$ holds. Therefore dealing with time scales $t\gg 1$ only the dependence of the denominator on the argument $s$ should be taken into account in evaluating Exps.~\eqref{sec2:GabL}. The other terms may be calculated within the limit $s\to 0$. In this way, using formula~\eqref{AppF:Unit2b} derived in \ref{AppF}, the Green function components $G^{\vartriangle,\triangledown}(\eta,s)$ determined by Exps.~\eqref{sec2:GabL} can be represented as 
\begin{equation}\label{sec2:New1}
G^{\vartriangle,\triangledown}(\eta,s) = \left[ 
                       s  - \left.\frac{dp^\text{st}(\eta')}{d\eta'}\right|_{\eta'=\theta}\right]^{-1} 
                      \frac{\alpha G^\text{out,\,in}(\eta,s)\big|_{s\to 0}}{\left[e^{\alpha\Phi(\eta_u)}-e^{\alpha\Phi(\eta_l)}\right]\int_0^{\infty}d\eta' e^{-\alpha\Phi(\eta')}}
\end{equation}
in the leading order. Then appealing to Exp.~\eqref{AppG:GGFinal2} the desired Green function or, speaking more strictly, its Laplace transform $G(\eta,s)$ specified by Exp.~\eqref{sec2:G0} is written as 
\begin{equation}\label{sec2:New2}
G(\eta,s) = \left[ s  - \left.\frac{dp^\text{st}(\eta')}{d\eta'}\right|_{\eta'=\theta}\right]^{-1}  p^\text{st}(\eta)\,.
\end{equation}
It should be reminded that here $p^\text{st}(\eta)$ is the stationary distribution function of the random variable $\eta$ after merging the half-spaces $\{\eta >0\}$ and $\{\eta <0\}$. 

Expression~\eqref{sec2:New2} for the Green function together with the constructed probabilistic weight $\mathfrak{P}_k(t)$ of the peak pattern $\mathbb{P}(t|k)$, Exps.~\eqref{sec2:G2} and \eqref{sec2:G2L}, are the main results of the given section. They enable us to consider the probabilistic properties of individual peaks in the pattern $\mathbb{P}(t|k)$ instead of random trajectories as whole entities, which is the subject of the next section.

\section{Peak pattern statistics}

As follows from the governing equation~\eqref{sec1:eq1} fluctuations of the particle velocity $v$ are correlated on time scales about unity, $t\sim 1$. Therefore, in the analysis of L\'evy flights based on model~\eqref{sec1:eq1} within the limit $t\gg 1$ the information about the particle velocity $\eta$ (or $v$ in the initial units) is redundant. So to rid our constructions of such details let us integrate Exp.~\eqref{sec2:New2} over all the possible values of the velocity $\eta$, i.e. over the region $0<\eta<\theta$. Since in this context the influence of the upper boundary $\theta \gg 1$ is ignorable, the integration region may be extended over the whole interval $0<\eta<\infty$. Then denoting the result of this action on the left-hand side of Exp.~\eqref{sec2:New2} by 
\begin{equation}\label{sec22:0}
   \mathfrak{G}_L(s) = \int\limits_0^\infty G(\eta,s)\,d\eta
\end{equation}
we get 
\begin{equation}\label{sec22:1}
    \mathfrak{G}_L(s) = \left[ s  - \left.\frac{dp^\text{st}(\eta)}{d\eta}\right|_{\eta=\theta}\right]^{-1} 
    = \tau	\sum_{k= 0}^\infty \left[F^\text{in}(s)F^\text{out}(s)\right]^k.
\end{equation}
In deriving the second equality of Exp.~\eqref{sec22:1} formula~\eqref{AppF:Unit2b} for the probabilistic weight of the composed unit $\mathbb{F}_{i,i+1}^\text{out}\otimes\mathbb{F}_{i+1,i+2}^\text{in}$ of the peak pattern $\mathbb{P}(t\mid k)$ has been used. Namely, first, it has enabled us to write
\begin{equation}\label{sec22:2}
    F^\text{in}(s)F^\text{out}(s) = 1 - \tau \left[ s  - \left.\frac{dp^\text{st}(\eta)}{d\eta}\right|_{\eta=\theta}\right]\,,
\end{equation} 
where the introduced time scale 
\begin{equation}\label{sec22:3}
     \tau = \frac1{\alpha}\left[{e^{\alpha\Phi(\eta_u)}-e^{\alpha\Phi(\eta_l)}}\right] 
           \int\limits_{0}^\infty d\eta'\, e^{-\alpha\Phi(\eta')}
\end{equation}
can be interpreted as the mean duration of this unit. Finally, the equality
\begin{equation*}
\sum_{k=0}^\infty \left[F^\text{in}(s)F^\text{out}(s) \right]^k = \frac1{1-F^\text{in}(s)F^\text{out}(s)}
\end{equation*}
has led us to Exp.~\eqref{sec22:1}. 

Since the limit of large time scales, $t\gg 1$, is under consideration, the inequalities  $s\ll 1$ and $\theta \gg 1$ are assumed to hold beforehand. In this case, by virtue of Exp.~\eqref{sec22:2},
\begin{equation*}
1-F^\text{in}(s)F^\text{out}(s)\ll 1
\end{equation*}
and in sum~\eqref{sec22:1} the terms with $k\gg 1$ contribute mainly to its value. Thereby we may confine our consideration to peak patterns $\mathbb{P}(t\mid k)$ that are composed of many elementary units.  It allows us to convert the discrete sum $\sum^\infty_{k=0}(\ldots)$ into a continuous integral as follows
\begin{equation}\label{sec22:4}
\sum_{k=0}^\infty (\ldots)\Rightarrow \int\limits_0^\infty dk(\ldots)
\end{equation}
and make use of the approximation
\begin{equation}\label{sec22:5}
\left[F^\text{in}(s)F^\text{out}(s)\right]^k = \exp\left\{-k \tau \left[ s  - \left.\frac{dp^\text{st}(\eta)}{d\eta}\right|_{\eta=\theta}\right]\right\}\,.
\end{equation}
In these terms formula~\eqref{sec22:1} reads
\begin{equation}\label{sec22:6}
	\mathfrak{G}_L(s) = \tau\int\limits_0^\infty dk \exp\left\{-k \tau \left[ s  - \left.\frac{dp^\text{st}(\eta)}{d\eta}\right|_{\eta=\theta}\right]\right\}
   {} = \tau\int\limits_0^\infty dt\int\limits_0^\infty dk\, e^{-st} \delta(t-k\tau) 
      \exp\left\{k \tau \left.\frac{dp^\text{st}(\eta)}{d\eta}\right|_{\eta=\theta}\right\}.
\end{equation}
The last expression enables us to represent the original function $\mathfrak{G}(t)$ of the Laplace transform $\mathfrak{G}_L(s)$ as
\begin{equation}\label{sec22:7}
	\mathfrak{G}(t) = \tau\int\limits_0^\infty dk\, \delta(t-k\tau) 
      \exp\left\{k \tau \left.\frac{dp^\text{st}(\eta)}{d\eta}\right|_{\eta=\theta}\right\}
      =\exp\left\{t \left.\frac{dp^\text{st}(\eta)}{d\eta}\right|_{\eta=\theta}\right\}.
\end{equation}
In other words, within the given limit it is possible to consider that only the pattern $\mathbb{P}(t\mid k)$ with $k=t/\tau$ (or more strictly $k=[t/\tau]$) contributes to the function $\mathfrak{G}(t)$. 

The latter statement is the pivot point in our constructions that admits another interpretation of the function $\mathfrak{G}(t)$ and its efficient application in the theory of L\'evy flights. 
Appealing to approximation~\eqref{sec22:5} let us rewrite Exp.~\eqref{sec22:7} in the form
\begin{equation}\label{sec22:8}
  \mathfrak{G}(t)= \left[F^\text{in}(0)F^\text{out}(0)\right]^{k_t}\qquad\text{with}\qquad k_t=\frac{t}{\tau}\,.
\end{equation}
In what follows we will not distinguish between the ratio $t/\tau$ and the derived integer $k_t=[t/\tau]$ because for $k\gg 1$ their difference is of minor importance and will keep in mind this integer where appropriate in using the ratio $t/\tau$. Therefore formula~\eqref{sec22:8} can be read as 
\begin{subequations}\label{sec22:12}
\begin{align}
\label{sec22:9}
  \mathfrak{G}(t) & = \int\limits_0^\infty \mathfrak{P}_{k_t}(t')\,dt'\qquad\text{for $t\gg\tau\sim 1$}
\\
\intertext{or, by virtue of \eqref{sec2:G2},}
\label{sec22:12a}
   \mathfrak{G}(t) &= \idotsint\limits_{0<t_1<t_2 \cdots < t_{2k} < \infty}  dt_1dt_2\ldots dt_{2k} 
                              \prod_{i=0}^{k_t-1} \mathcal{F}^\text{in}(t_{2i+2}-t_{2i+1}) \mathcal{F}^\text{out}(t_{2i+1}-t_{2i})\,.
\end{align}
\end{subequations}
Expressions~\eqref{sec22:12} are one of the main technical results obtained in the present work. In particular, they allow us to state the following.

\begin{prop}\label{PropAdd:1}
The probability  $\mathfrak{G}(t)$ of finding the particle whose ``velocity'' $\eta$ is localized inside the region $|\eta| <\theta$ during the time interval $(0,t)$ for $\theta\,,t\gg 1$ is equal to the probability  $\mathfrak{P}_{k_t}(t')$ of the pattern $\mathbb{P}(t'|k_t)$ with $k_t = t/\tau$ peaks after the averaging over its duration $t'$.
\end{prop}

\noindent It should be noted that the given statement has been directly justified assuming the initial particle velocity to be equal to $\eta_0 = \eta_u$. However, as can be demonstrated, it holds for any $\eta_0\ll\theta$. Also it is worthwhile to underline the fact that within this description the physical time $t$ is replaced by the number $k_t$ of peaks forming the pattern  $\mathbb{P}(t'|k_t)$, whereas its actual duration $t'$ does not matter. It becomes possible due to a certain self-averaging effect. In particular, according to Exp.~\eqref{sec22:2}, the duration of the units  $\{\mathbb{F}_{i,i+1}^\text{out}\otimes\mathbb{F}_{i+1,i+2}^\text{in}\}$ should be distributed with the exponential law, at least, on scales $t\gg \tau$. As a result the duration $t'$ of the pattern  $\mathbb{P}(t'|k_t)$ changes near its mean value $k_t\tau=t$ with relatively small amplitude.
Besides, to simplify further explanations, when appropriate the probability  $\mathfrak{G}(t)$ will be also referred to as the probability of the pattern $\mathbb{P}(k_t)$, where the argument $t'$, its duration, is omitted to underline that the required averaging over $t'$ has been performed. 

Expression~\eqref{sec22:12a} admits another interpretation of the developed random walk classification.  

\begin{prop}\label{Prop1}
The probability $\mathfrak{G}(t)$ is equal to the probability of the particle returning to the initial point $\eta = \eta_u$ for $k_t=t/\tau$ times sometime after. The return events are understood in the sense determined by the sequence of alternate jumps between the two boundaries $\{\eta_l,\eta_u\}$ of the layer $\mathcal{L}$. 
\end{prop}

The following proposition will play a significant role in further constructions. 

\begin{prop}\label{Prop2}
The probability $\mathfrak{G}(t)$ of the peak pattern $\mathbb{P}(k_t)$ is specified by the expression
\begin{equation}\label{Prop.eq.1}
	\mathfrak{G}(t) = \exp\left\{t \left.\frac{dp^\text{st}(\eta)}{d\eta}\right|_{\eta=\theta}\right\}
      \end{equation}
and the probability of its one unit, i.e., the probability of one return to the initial point  $\eta = \eta_u$ sometime after is evaluated by the expression  
\begin{equation}\label{Prop.eq.2}
	\mathfrak{g}(t) = 1+\tau\left. \frac{dp^\text{st}(\eta)}{d\eta}\right|_{\eta=\theta}\,.
      \end{equation}
\end{prop}

\noindent
Formula~\eqref{Prop.eq.1} stems directly from Exp.~\eqref{sec22:6}, whereas formula~\eqref{Prop.eq.2} is a consequence of Exps.~\eqref{sec22:8} and \eqref{sec22:2}, where in the latter one the parameter $s$ is set equal to zero, $s=0$.

\section{Core stochastic process}

As noted above spatial displacement of the wandering particle is manly caused by extreme fluctuations in its velocity. Therefore the direct contribution of the random walks inside the layer $\mathcal{L}$ to the particle displacement is ignorable. We have to take into account only the fact that during a certain time the wandering particle or, more strictly, its velocity $\eta$ (or $v$ in the original units) is located inside this layer. Thus, if two given stochastic processes differ from each other in their properties \textit{only inside} the layer $\mathcal{L}$ but are characterized by the same probabilistic weight $\mathcal{F}^\text{in}(t)$ of the fragments $\{\mathbb{F}^\text{in}_{i,i+1}\}$, then they can be regarded as equivalent. In the case under consideration, i.e., for $t\gg 1$ it is sufficient to impose the latter requirement on the Laplace transform $F^\text{in}(s)$ of this weight within the linear approximation in $s\ll 1$. 

The purpose of the present section is to construct a certain stochastic process that, on one hand, is equivalent in this sense to the process governed by model~\eqref{sec1:eq1}. On the other hand, its description should admit an efficient scaling of the corresponding governing equation, which will enable us to single out the basic mechanism responsible for the found basic features of particle motion. The other characteristics of particle motion can be explained by appealing to the corresponding coefficients of this scaling. The desired equivalent process will be called the core stochastic process for these reasons. 

\subsection{Construction}

\begin{figure}
\begin{center}
\includegraphics[width=0.6\columnwidth]{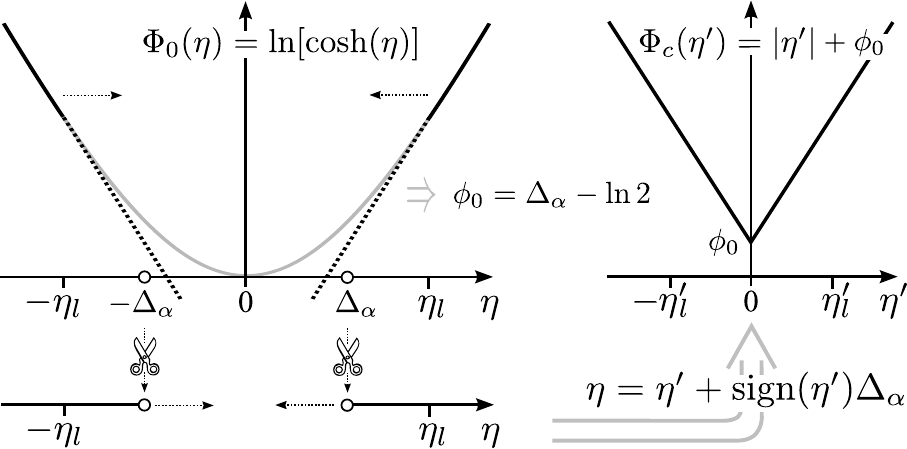}
\end{center}
\caption{Illustration of constructing an effective stochastic process $\{\eta'(t)\}$ with the V-type potential that is equivalent to the initial process $\{\eta(t)\}$ from the standpoint of the particle displacement in the space $\mathbb{R}_x$.}
\label{F5}
\end{figure}

\begin{figure}
\begin{center}
\includegraphics[width=0.45\columnwidth]{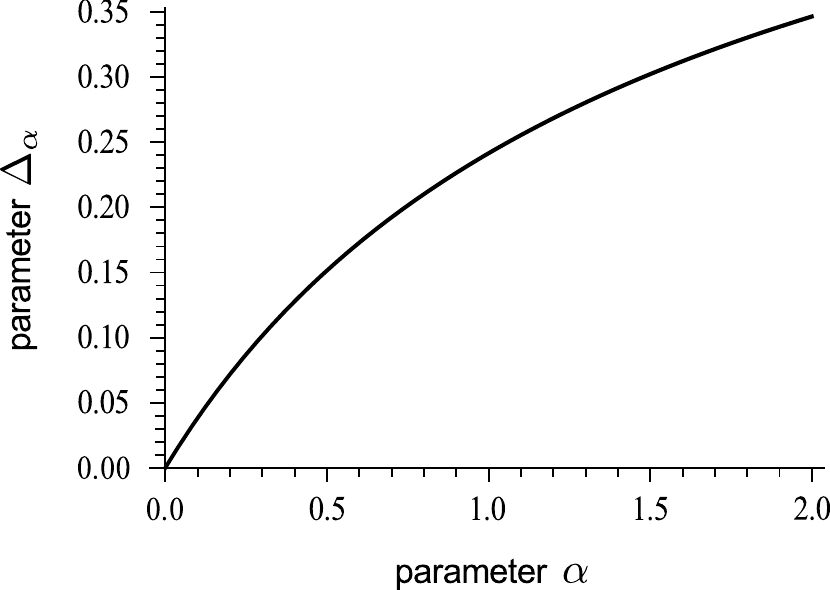}
\end{center}
\caption{The magnitude of the parameter $\Delta_\alpha$ vs the possible values of the parameter $\alpha$.}
\label{F6}
\end{figure}

The key points in constructing the  core stochastic process  are illustrated in Fig.~\ref{F5}. The construction is based on the transformation 
\begin{align}
\label{sec3:1}
	\eta & = \eta'+ \sign(\eta') \Delta_\alpha
\\
\intertext{and the replacement of the potential $\Phi_0(\eta)$ by its linear extrapolation to the region $\eta\lesssim 1$}
\label{sec3:2}
  \Phi_0(\eta) & \rightarrow |\eta'| + \Delta_\alpha - \ln2\,.
\end{align}
The constant $\Delta_\alpha$ has to be chosen such that in both the cases the function $F^\text{in}(s)$ take the same form within the linear approximation in $s$. Appealing to \ref{AppFIN}, namely, formula~\eqref{AppF:Fin} we see that this requirement is reduced practically to the equality 
\begin{equation}\label{sec3:3}
	\int\limits_0^\infty e^{-\alpha\Phi_0(\eta)} d\eta = \int\limits_{\Delta_\alpha}^\infty e^{-\alpha(\eta-\ln2)}d\eta
\end{equation}
because the potential $\Phi_0(\eta)$ differs considerably from its linear interpolation~\eqref{sec3:2} only in the region $\eta \lesssim 1$. The direct calculation of the integrals entering equality~\eqref{sec3:3} for function~\eqref{sec1:Phi0} yields the desired value
\begin{equation}\label{sec3:4}
  \Delta_\alpha = \frac1{\alpha}\ln\left[
   \frac{4\Gamma(\alpha)}{\alpha\Gamma^2(\alpha/2)} 
  \right]\,,
\end{equation}
where $\Gamma(\ldots)$ is the gamma function. Figure~\ref{F6} shows the value $\Delta_\alpha$ as a function of the parameter $\alpha\in(0,2)$. It should be noted that previously the value of $\Delta_\alpha$ has been implicitly assumed to be less than $\eta_l$, which is justified by Fig.~\ref{F6}. Indeed, on one side, the boundary $\eta_l$ is initially assumed to belong to the region wherein the function $\Phi_0(\eta)$ admits the linear approximation, i.e., the estimate $\eta_l\gtrsim1$ is assumed to hold beforehand. On the other side, the maximal value of $\Delta_\alpha$ is about 0.35 according to  Fig.~\ref{F6}.

In what follows we will confine our consideration to the special case~\eqref{sec1:kg0} matching the ideal L\'evy flights. It enables us to convert from the stochastic process $\{\eta(t)\}$ to the effective stochastic process $\{\eta'(t)\}$ using transformation~\eqref{sec3:1} and to reduce Eq.~\eqref{sec1:eq2} to the stochastic equation
\begin{equation}\label{sec3:5}
  \frac{d\eta'}{dt}=-\alpha\sign(\eta')+\sqrt{2}\xi(t)
\end{equation}
governing the analyzed L\'evy flights in the equivalent way.
Then, via the next transformation $\eta'\to \mathfrak{u}$ specified by the expressions
\begin{equation}\label{sec3:6}
   \eta' = \frac1{\alpha}\,\mathfrak{u}\,,\qquad t = \frac1{\alpha^2}\,\mathfrak{t}
\end{equation}
involving also the transformation of the time scales $t\to \mathfrak{t}$ equation~\eqref{sec3:5} is rewritten as follows
\begin{equation}\label{sec3:7}
  \frac{d\mathfrak{u}}{d\mathfrak{t}}=-\sign(\mathfrak{u})+\sqrt{2}\xi(\mathfrak{t})\,.
\end{equation}
It is a parameter-free stochastic differential equation with additive white noise such that 
\begin{equation}\label{sec3:8}
  \left< \xi(\mathfrak{t})\right> = 0\,, \qquad \left< \xi(\mathfrak{t}) \xi(\mathfrak{t}')\right> = \delta(\mathfrak{t}-\mathfrak{t}')\,. 
\end{equation}
In other words, we have constructed the stochastic process $\{\mathfrak{u}(\mathfrak{t})\}$ governed by Eq.~\eqref{sec3:7} that can be treated as the desired \textit{core process}. It is of the same form for all the types of one-dimensional L\'evy flights, at least, L\'evy flights described by models similar to Eq.~\eqref{sec1:eq1} inside the region where the cut-off effects are not significant. The basic parameters characterizing the generated L\'evy flights such as the exponent of the L\'evy scaling law depend on the system parameters, in particular, the coefficient $\alpha$ via the transformation from  $\mathfrak{u}(\mathfrak{t})$ to $v(t)=dx/dt$. Combining together Exps.~\eqref{sec1:voneta0}, \eqref{sec3:1}, and \eqref{sec3:6} we can write this transformation as follows
\begin{equation}\label{sec3:9}
  \frac{dx}{dt} = \sinh\left\{
  \frac{\mathfrak{u}(\mathfrak{t})}{\alpha} + \Delta_\alpha \sign\left[\mathfrak{u}(\mathfrak{t})\right]
  \right\}
\end{equation}
which is completed by the second proportionality of Exps.~\eqref{sec3:6} relating $t$ to $\mathfrak{t}$.

Summarizing aforesaid we get the following statement.
\begin{prop}\label{Prop3}
The L\'evy flights $\{x(t)\}$ governed by model~\eqref{sec1:eq1} with the kinetic coefficients~\eqref{sec1:kg0} are equivalently described by the parameterless stochastic process $\{\mathfrak{u}(\mathfrak{t})\}$ obeying the stochastic differential equation~\eqref{sec3:7} with additive white noise.  The variable $\mathfrak{u}$ and the initial variable $x$ are related via Exp.~\eqref{sec3:9} and the second proportionality of Exps.~\eqref{sec3:6}. 
\end{prop}

Propositions~\ref{PropAdd:1}--\ref{Prop3} are the basic results of the present paper. In what follows we will make use of them to demonstrate the fact that model~\eqref{sec1:eq1} does describe the L\'evy flights of superballistic, quasiballistic, and superdiffusive regimes matching $0<\alpha < 1$, $\alpha = 1$, and $1<\alpha <2$, respectively. It should be reminded that this correspondence  was  strictly proved only for the superdiffusive L\'evy flights \cite{we2} and for the other regimes it was demonstrated numerically \cite{we3}. A more sophisticated analysis of such L\'evy flights based on the constructed representation is worthy of an individual publication.

\subsection{Asymptotic properties of the core stochastic process}

Let us analyze the probabilistic properties of the peak pattern $\mathbb{P}_\mathfrak{u}(k_\mathfrak{t})$ for the core stochastic process $\mathfrak{u}(\mathfrak{t})$ in the limit $\theta\gg1$. To do this, the asymptotic behavior of distribution functions describing the statistics of time patterns related to  $\mathbb{P}_\mathfrak{u}(k_\mathfrak{t})$  as $\theta\to\infty$ will be considered in detail. Dealing with the core stochastic process we may set the first boundary $\mathfrak{u}_l$ of the layer $\mathcal{L}_\mathfrak{u}$ equal to zero, $\mathfrak{u}_l=0$, without loss of generality.  The stationary distribution function $\mathfrak{p}^\text{st}(\mathfrak{u})$ of the random variable $\mathfrak{u}$  is determined by the expression
\begin{equation}\label{sec3:10}
	\mathfrak{p}^\text{st}(\mathfrak{u}) = e^{-\mathfrak{u}}
\end{equation}
after merging the half-spaces $\{\mathfrak{u}<0\}$ and $\{\mathfrak{u}>0\}$, which stems directly from Exps.~\eqref{sec1:Phi} and \eqref{petast} after setting $\phi(\eta) = 1$ and $\alpha=1$ in these formulas. 

The probability $\mathfrak{G_u}(\mathfrak{t},\theta)$ is determined by the corresponding path integral over all the trajectories $\{\mathfrak{u}(\mathfrak{t}')\}_{0}^{\mathfrak{t}}$ meeting the inequality $0<\mathfrak{u}(\mathfrak{t}')< \theta$ for $0<\mathfrak{t}'<\mathfrak{t}$; now the value $\theta$ is noted directly in the list of the arguments of the function  $\mathfrak{G_u}(\mathfrak{t},\theta)$. Therefore the function
\begin{equation}\label{sec3:11}
\mathfrak{F}(\mathfrak{t},\theta) := \frac{d\mathfrak{G}(\mathfrak{t},\theta)}{d\theta} = \mathfrak{t}e^{-\theta}\cdot \exp\left\{-\mathfrak{t}e^{-\theta}\right\}
\end{equation}
is the probability density that the random variable $\mathfrak{u}$ attains the maximal value equal to $\theta$, i.e.,
\begin{equation}\label{sec3:12}
  \theta = \max_{0<\mathfrak{t}'<\mathfrak{t}}\mathfrak{u}(\mathfrak{t}')
\end{equation}
inside the interval $(0,\mathfrak{t})$. The latter equality in Exp.~\eqref{sec3:11} stems directly from Proposition~\ref{Prop2} and formula~\eqref{sec3:10}. According to Proposition~\ref{PropAdd:1} the function  $\mathfrak{G_u}(\mathfrak{t},\theta)$ admits the interpretation in terms of the peak pattern  $\mathbb{P}_\mathfrak{u}(k_\mathfrak{t})$, where the number of peaks or, what is the same, the number of basic units $k_\mathfrak{t} := k_t =\mathfrak{t}/(\alpha^2\tau)$ is regarded as a fixed parameter of the random walk classification. In contrast, the specific time moments $\{\mathfrak{t}_i\}$ of the pattern partition including the terminal point $\mathfrak{t}_n = \mathfrak{t}$ (Fig.~\ref{F3}) are treated as internal classification parameters over which the averaging must be performed. Under these conditions the function $\mathfrak{G_u}(\mathfrak{t},\theta)$ is just the the probability of the peak pattern $\mathbb{P}_\mathfrak{u}(k_\mathfrak{t})$ generated by the random variable $\mathfrak{u}$ and containing exactly $k_\mathfrak{t}$ peaks. Therefore the function $\mathfrak{F}(\mathfrak{t},\theta)$ introduced via Exp.~\eqref{sec3:11} gives the probability density that the maximum attained by the variable $\mathfrak{u}$  within at least one of the $k_\mathfrak{t}$ peaks is equal to $\theta$. Naturally, the maxima attained by this variable  inside the other peaks must be less than or equal to $\theta$.
Going in a similar way the probability density $\mathfrak{f}(\theta)$ of the extreme value $\theta$ for one peak is written as 
\begin{equation}\label{sec3:13}
\mathfrak{f}(\theta) := \frac{d\mathfrak{g}(\theta)}{d\theta} =\tau_\mathfrak{u} e^{-\theta}\,.
\end{equation}
Here the value $\theta$ has the meaning
\begin{equation}\label{sec3:14}
  \theta = \max_{\mathfrak{t}'\in\text{a given peak}}\mathfrak{u}(\mathfrak{t}')
\end{equation}
and, as before, $\tau_\mathfrak{u}$ is the mean duration of individual peaks. Function~\eqref{sec3:11} is plotted in Fig.~\ref{F7} and its asymptotic behavior as $\theta\to \infty$ enables us to introduce a single-peak approximation.

\subsubsection{Single-peak approximation}

\begin{figure}
\begin{center}
\includegraphics[width=0.5\columnwidth]{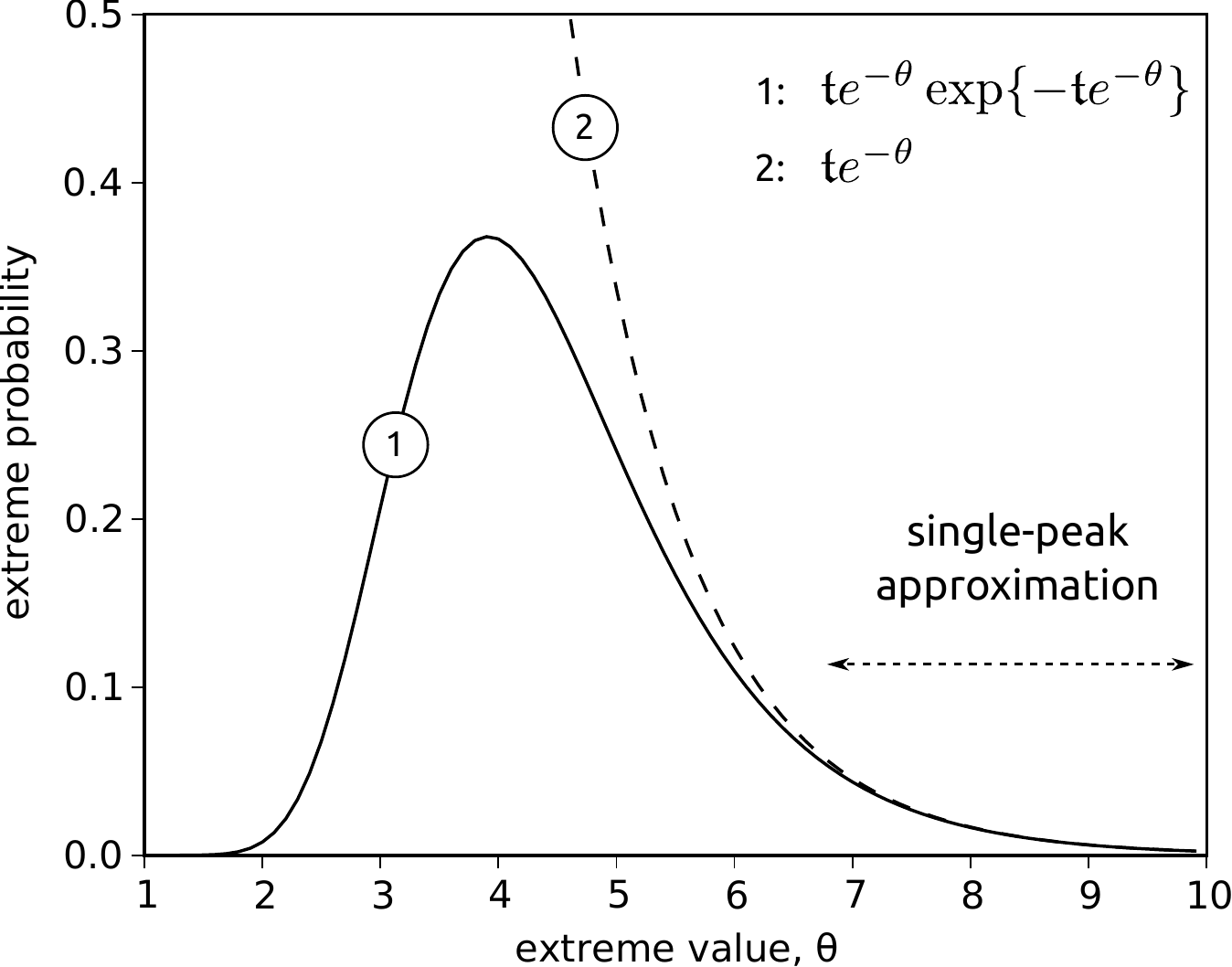}
\end{center}
\caption{Probability density $\mathfrak{F}(\theta,\mathfrak{t})$ of the extreme value $\theta$ being attained by the random variable $\mathfrak{u}$ within the peak pattern  $\mathbb{P}_\mathfrak{u}(k_\mathfrak{t})$ (curve 1) and its form in the frameworks of the single-peak approximation (curve 2). In plotting these data the time $\mathfrak{t} =50$ was used.}
\label{F7}
\end{figure}

In the region $\theta \gtrsim \theta_\mathfrak{t}$, where $\mathfrak{t}>0$ and
\begin{equation}\label{sec3:15}
	\theta_\mathfrak{t} = \ln(\mathfrak{t})
\end{equation}
so that $\mathfrak{t}e^{-\theta_\mathfrak{t}} = 1$, the asymptotic behavior of the extreme value probability $\mathfrak{F}(\theta,\mathfrak{t})$ is specified by the expression 
\begin{equation}\label{sec3:15.1}
	\mathfrak{F}(\theta,\mathfrak{t})\approx \mathfrak{t}e^{-\theta}
\end{equation}
by virtue of \eqref{sec3:11}. Formula~\eqref{sec3:15.1} can be reproduced in the following way. Let us assume, at first, that the maximal value $\theta$ is attained inside a certain peak $i$. Then inside the other $(k_\mathfrak{t}-1)$ peaks the random variable 
$\mathfrak{u}$ has to belong to the interval $\mathfrak{u}\in(0,\theta)$. The probability of the first event is given by the function $\mathfrak{f}(\theta)$. The probability of the second event can be written as 
\begin{equation*} 
 \mathfrak{g}(\theta)^{k_\mathfrak{t}-1}\approx \left[1 -\mathfrak{f}(\theta)\right]^{k_\mathfrak{t}}\,.
\end{equation*}
In deriving this formula Exps.~\eqref{Prop.eq.2}, \eqref{sec3:10}, and \eqref{sec3:13} have been used as well as the inequality $k_\mathfrak{t} = \mathfrak{t}/\tau_\mathfrak{u}\gg 1$ has been taken into account. The latter inequality enables us to confine out consideration to the patten configurations for which $\mathfrak{f}(\theta)\ll 1$ and, thus, to ignore the difference between $k_\mathfrak{t}$ and $(k_\mathfrak{t}-1)$. Therefore, the probability of the extreme value $\theta$ being attained inside peak $i$ is 
\begin{equation*} 
  \mathfrak{f}(\theta)\left[1 -\mathfrak{f}(\theta)\right]^{k_\mathfrak{t}}\approx 
  \mathfrak{f}(\theta)\exp\left\{-k_\mathfrak{t} \mathfrak{f}(\theta)\right\}\,.
\end{equation*}
The choice of peak $i$ is arbitrary. So to find the probability of the extreme value $\theta$ being attained somewhere inside the peak pattern $\mathbb{P}_\mathfrak{u}(k_\mathfrak{t})$ we have to multiply the last expression by the number $k_\mathfrak{t} = \mathfrak{t}/\tau_\mathfrak{u}$ of peaks, which gives rise to the equality
\begin{equation} \label{sec3:18}
 	\mathfrak{F}(\theta,\mathfrak{t}) =   k_\mathfrak{t} \mathfrak{f}(\theta) \exp\left\{-k_\mathfrak{t} \mathfrak{f}(\theta)\right\}
\end{equation}
coinciding with Exp.~\eqref{sec3:11} by virtue of \eqref{sec3:13}.
In the case $\theta \gtrsim \theta_\mathfrak{t}$ the inequality $k_\mathfrak{t}\mathfrak{f}(\theta)\ll 1$ holds and the exponential multiplier in Exp.~\eqref{sec3:18} can be ignored, leading to Exp.~\eqref{sec3:15.1}. In other words, the extreme value $\theta$ is attained actually within one peak whereas the effect of the other peaks on its probabilistic properties is ignorable.

Summarizing these arguments we draw the conclusion below. 

\begin{prop}[The single-peak approximation]\label{Prop4}

In the region $\theta \gtrsim \theta_\mathfrak{t}$ the probability density $\mathfrak{F}(\theta,\mathfrak{t})$ of the variable $\mathfrak{u}$ attaining the extremal value $\theta$  within the peak pattern  $\mathbb{P}_\mathfrak{u}(k_\mathfrak{t})$ exhibits the asymptotic behavior that can be represented in the following way. The value $\theta$ is attained exactly within one peak. Variations of the random variable $\mathfrak{u}$ inside the other peaks may be assumed to be small in comparison with $\theta$ and the condition $\mathfrak{u}\leq \theta$ does not really affect their probabilistic properties.

\end{prop} 

\noindent Naturally, when a given value of $\theta$ falls outside the region  $\theta \gtrsim \theta_\mathfrak{t}$, i.e., $\theta \lesssim \theta_\mathfrak{t}$ variations of the random variable $\mathfrak{u}$ in many peaks become comparable  with $\theta$. As a result, the fact that the condition $\mathfrak{u}<\theta$ is imposed on all the peaks of $\mathbb{P}_\mathfrak{u}(k_\mathfrak{t})$ is essential and the single-peak approximation does not hold. In particular, for  $\theta \lesssim \theta_\mathfrak{t}$ the probability density $\mathfrak{F}(\theta,\mathfrak{t})$ has to deviate substantially from its asymptotics~\eqref{sec3:15.1} as illustrated in Fig.~\ref{F7}. 

Now let us apply these constructions to the analysis of the particle displacement in the space $\mathbb{R}_x$.

\section{Asymptotic properties and scaling law of spatial particle displacement}

The purpose of this section is to demonstrate the fact that for a particle whose random motion is governed by model~\eqref{sec1:eq1} the asymptotic behavior of the distribution function $\mathcal{P}(x,t)$ of its spatial displacement $x$ during the time interval $t$ is determined by the asymptotic properties of the peak pattern  $\mathbb{P}_\mathfrak{u}(k_\mathfrak{t})$. By virtue of Proposition~\ref{Prop4} it can be reformulated as follows.

\begin{prop}\label{Prop5}
Large fluctuations in the spatial displacement $x$ of the wandering particle during the time interval $t\gg1$ are related mainly to its motion inside the single peak of the pattern $\mathbb{P}_\mathfrak{u}(k_\mathfrak{t})$ whose  amplitude is maximal in comparison with the other peaks and attains extremely large values.
\end{prop}

Proposition~\ref{Prop5} is actually the implementation of the single-peak approximation applied to describing the particle motion in the space $\mathbb{R}_x$. Appealing to the notion of the core stochastic process $\{\mathfrak{u(t)}\}$ we also can formulate the next statements.

\begin{prop}\label{Prop7}
The region $\mathbb{L}_x(t)=\{x : |x|\gg \overline{x}(t)\}$ of large fluctuations in the particle spatial displacement $x$ directly matches the region $\theta\gtrsim\theta_\mathfrak{t}$ of the extreme values $\theta$ attained by the random variable $\mathfrak{u(t)}$ inside the peak pattern $\mathbb{P}_\mathfrak{u}(k_\mathfrak{t})$.
\end{prop}

\begin{prop}\label{Prop8}
The asymptotic behavior of the distribution function $\mathcal{P}(x,t)$ in the region $\mathbb{L}_x(t)$ is practically specified by the asymptotic behavior of the probability density $\mathfrak{F}(\theta,\mathfrak{t})$ inside the region  $\theta\gtrsim\theta_\mathfrak{t}$.
\end{prop}

\begin{prop}\label{Prop9}
The lower boundary $\overline{x}(t)$ of the region $\mathbb{L}_x(t)$ quantifying the characteristic displacement of the wandering particle during the time interval $t$ can be evaluated using the single-peak approximation and setting $\theta$ equal to the lower boundary of the $\theta$-region wherein the single-peak approximation holds, $\theta\sim \theta_\mathfrak{t}$.
\end{prop}

Below in this section, at first, we will accept Propositions~\ref{Prop5}--\ref{Prop9} as hypotheses and using them construct the distribution function $\mathcal{P}(x,t)$. Then the comparison  of the results to be obtained below with the previous rigorous results \cite{we2,we3} justifies these Propositions.  

Integrating expression~\eqref{sec3:9} over the time interval $(0,t)$ we get the relationship between the particle spatial displacement $x(t)$ during the given time interval and the time pattern $\{\mathfrak{u(t')}\}_{\mathfrak{t'}=0}^{\mathfrak{t'=t}}$ of the core stochastic process
\begin{equation}\label{new:1}
  x(t) = \frac1{\alpha^2}\int\limits_{0}^{\mathfrak{t}}\sinh\left\{
    \frac{\mathfrak{u}(\mathfrak{t'})}{\alpha} + \Delta_\alpha \sign\left[\mathfrak{u}(\mathfrak{t'})\right]
    \right\} d\mathfrak{t'}\,.
\end{equation}
Here the time scale transformation~\eqref{sec3:6} has been taken into account. According to Proposition~\ref{Prop5} the time integration in Exp.~\eqref{new:1} can be reduced to the integration over the largest peak $\mathbb{P}_\theta$ of the pattern $\{\mathfrak{u(t')}\}_{\mathfrak{t'}=0}^{\mathfrak{t'=t}}$ provided extreme fluctuations of the random variable $x(t)$ are under consideration. In addition, Proposition~\ref{Prop7} claims that the random variable $\mathfrak{u}$ inside the peak $\mathbb{P}_\theta$ attains values much larger than unity. The latter holds, at least, within a certain neighborhood of the maximum $\theta$ attained by the variable $\mathfrak{u}$ inside the peak $\mathbb{P}_\theta$. In this case formula~\eqref{new:1} can be rewritten as  
\begin{equation}\label{sec4:1}
	x(t) =\frac{e^{\Delta_\alpha}}{2\alpha^2} \int\limits_{\mathfrak{t}\in \mathbb{P}_\theta}\exp\left\{\frac{\mathfrak{u}(\mathfrak{t})}{\alpha}\right\}d\mathfrak{t}\,.
\end{equation}
In obtaining this expression we implicitly have assumed without loss of generality that the variable  $\mathfrak{u(t)}$ takes positive values inside the peak $\mathbb{P}_\theta$.

The spatial displacement $x$ of the wandering particle and its velocity maximum 
\begin{equation}\label{sec4:2}
\vartheta = \frac{e^{\Delta_\alpha}}2 \exp\left\{\frac{\mathfrak{\theta}}{\alpha}\right\}
\end{equation}
attained inside the peak $\mathbb{P}_\theta$ are partly independent variables. Indeed, for example, their ratio 
\begin{equation}\label{sec4:2a}
\frac{x}{\vartheta} = \frac1{\alpha^2}  \int_{\mathfrak{t}\in \mathbb{P}_\theta}\exp\left\{\frac{\mathfrak{u}(\mathfrak{t})-\theta}{\alpha}\right\}d\mathfrak{t}
\end{equation}
depends on the details of the pattern $\mathfrak{u}(\mathfrak{t})$ in the vicinity of its maximum $\theta$. Nevertheless, these details seem not to be too essential; they determine mainly some cofactors of order unity, see also Ref.~\cite{we1}.

\begin{figure}
\begin{center}
\includegraphics[width=0.5\columnwidth]{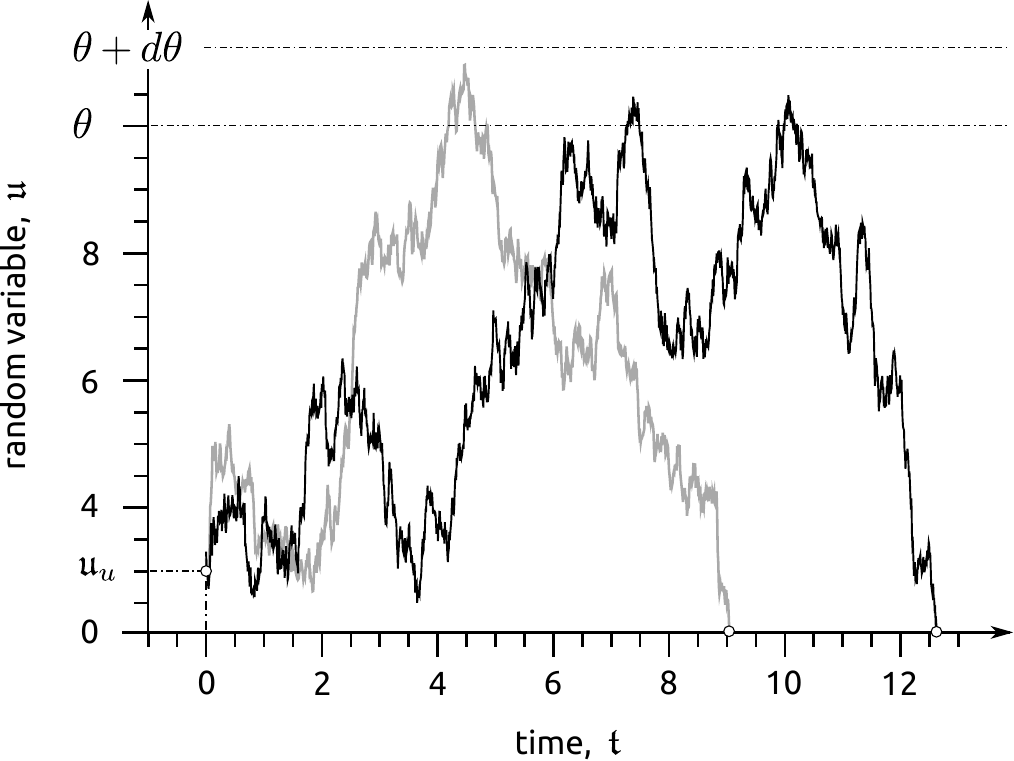}
\end{center}
\caption{Example of random trajectories $\{\mathfrak{u}(\mathfrak{t})\}$ implementing the peak $\mathbb{P}_\theta$. Initially, $\mathfrak{t}=0$, both of them start from the point $\mathfrak{u}_u$ and terminate when crossing the lower boundary $\mathfrak{u}_l = 0$ for the first time. This figure also illustrates the technique of constructing the distribution function of the maximal value $\theta$ attained by \textit{continuous} random walks $\{\mathfrak{u}(\mathfrak{t})\}$ via counting the trajectories passing through the layer $(\theta,\theta+d\theta)$ without touching its upper boundary $u = \theta+d\theta$. The division of the result by the thickness $d\theta$ of the analyzed layer yields the probability \textit{density} $\mathfrak{f}(\theta)$.}
\label{F8}
\end{figure}

\begin{figure}
\begin{center}
\includegraphics[width=0.9\columnwidth]{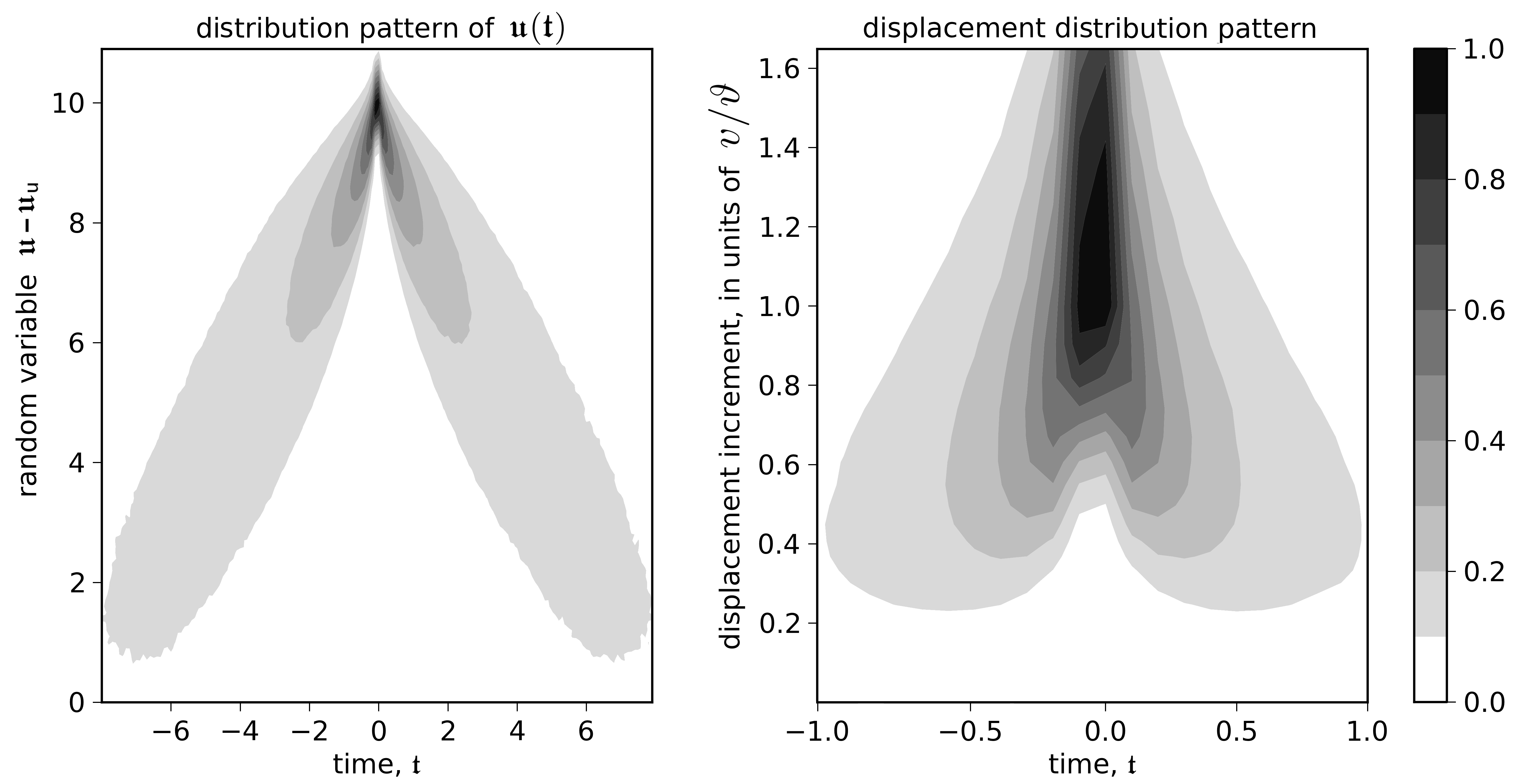}	
\end{center}
\caption{Spatial patterns visualizing the regular trend in the dynamics of the random variable $\mathfrak{u(t)}$ in the vicinity of the attained maximal value $\theta$ and the scattering of  $\mathfrak{u(t)}$ around the regular trend. The left window depicts this pattern on the plane $\{\mathfrak{t,u}\}$, the right window maps this pattern on the plane $\{\mathfrak{t},\delta x\}$. Here $\delta x$ is the elementary displacement of the wandering particle along the axis $x$ during the time interval $d\mathfrak{t}$ or, what is actually the same after the corresponding normalization, the particle velocity $v$ normalized to the velocity $\vartheta$ corresponding to $\mathfrak{u}=\theta$. In obtaining these data Eq.~\eqref{sec3:7} and Exp.~\eqref{sec3:9} with $\alpha = 1$ were used; the extreme value was set equal to $\theta = 10$. The details of constructing the given patterns are described in the text.}
\label{F9}
\end{figure}

To justify the latter statements, first, Figure~\ref{F8} depicts two trajectories $\mathfrak{u}(\mathfrak{t})$ implementing the peak $\mathbb{P}_\theta$. The difference in their forms explains the partial independence of the variable $x$ and $\vartheta$. Second, Figure~\ref{F9} demonstrates the statistical properties of such trajectories. The shown patterns were obtained in the following way. A collection of random trajectories similar to ones shown in Fig.~\ref{F8} were generated based on Eq.~\eqref{sec3:7} with the discretization time step of 0.01. All the trajectories started from the point $\mathfrak{u}_u = 1$ and terminated when crossing the boundary $\mathfrak{u}_l = 0$ for the first time. Only the trajectories that passed through the layer $(\theta,\theta+1)$ with $\theta = 10$ without touching the upper boundary $\theta+1 = 11$ were taken into account. Then for each trajectory the time moment $\mathfrak{t}_\text{max}$ of attaining the corresponding maximum $\mathfrak{u}_\text{max}$ was fixed and the trajectory as a whole was shifted along the time axis that the point $\mathfrak{t}_\text{max}$ be located at the time origin $\mathfrak{t}=0$. In this way all the trajectories were rearranged that their maxima be located at the same point on the time axis. The total number of the trajectories constructed in this way was equal to $10^5$. 
Then the plane $\{\mathfrak{t,u}\}$ was partitioned into cells of $0.1\times 0.1$ size and the discretization points of individual trajectories fell into each cell were counted. Finally their numbers were renormalized to the obtained maximum. The left window in Fig.~\ref{F9} exhibits the obtained distribution of these values called the distribution pattern of $\mathfrak{u(t)}$. Actually this pattern visualizes the regular trend in the dynamics of the variable $\mathfrak{u(t)}$ near the extreme point $\theta$ and its scattering around it. As should be expected, the regular trend of  $\mathfrak{u(t)}$ matches the optimal trajectory 
\begin{equation}\label{new:2}
  \mathfrak{u}_\text{opt}(\mathfrak{t}) = \theta - |\mathfrak{t-t}_\text{max}|
\end{equation}
of the system motion towards the maximum $\theta$ and away from it. The symmetry of this pattern is worthy of being noted because only the left branch of the optimal trajectory~\eqref{new:2} matching the motion towards the maximum ($\mathfrak{t< t}_\text{max}$) is related to extreme fluctuations in the time dynamics of $\mathfrak{u(t)}$. The right one is no more than a ``free'' motion of particle under the regular drift. The right window depicts actually the same pattern in units of the particle elementary displacement, see Exp.~\eqref{sec4:1}, 


\begin{equation*}
	\delta x \propto \exp \left\{\frac{\mathfrak{u}(\mathfrak{t})}{\alpha}\right\} d\mathfrak{t}
\end{equation*}
during the time step $d\mathfrak{t}$. It plots this pattern normalized to the particle velocity $\vartheta$ attained at $\mathfrak{u} = \theta$. As seen in this figure, fluctuations in the variable $x$ should be comparable with its mean value or less than it. In constructing the given patterns the value of $\alpha=1$ was used. Therefore, in spite of the partial independence of the random variables $x$ and $\vartheta$ solely the statistical properties of the variable $\vartheta$ are responsible for the L\'evy characteristics of the generated random walks. 

For the sake of simplicity in the present analysis we confine our consideration to the regular model of the time variations $\mathfrak{u(t)}$ near the extremum point, i.e., the ansatz
\begin{equation}\label{sec4:3}
 \mathfrak{u}(\mathfrak{t}) = \mathfrak{u}_\text{opt}(\mathfrak{t}) = \theta - |\mathfrak{t-t}_\text{max}|
\end{equation}    
will be used in calculating integral~\eqref{sec4:1}. An approach enabling us to go beyond this approximation will be published somewhere else. Substituting \eqref{sec4:3} into \eqref{sec4:1} we get 
\begin{equation}\label{sec4:4}
	x =\frac{e^{\Delta_\alpha}}{\alpha} \exp\left\{\frac{\theta}{\alpha}\right\}\,.
\end{equation}
Then using formula~\eqref{sec3:4} and Exp.~\eqref{sec3:15.1} for the extreme value probability density $\mathfrak{F}(\theta,\mathfrak{t})$ we obtain the expression
%
\begin{equation}\label{sec4:5}
   \mathcal{P}(x,t) = \frac{4\alpha\Gamma(1+\alpha)}{\alpha^\alpha \Gamma^2(\alpha/2)}\cdot \frac{t}{x^{1+\alpha}}
\end{equation}
giving us the asymptotics of the distribution function $\mathcal{P}(x,t)$ of the particle spatial displacement $x$ during the time interval $t$ when $x\gg \overline{x}(t)$ and the expression 
\begin{equation}\label{sec4:6}
   \overline{x}(t)=  \left[\frac{4\Gamma(1+\alpha)}{\alpha^\alpha \Gamma^2(\alpha/2)}\cdot t\,\right]^{1/\alpha}
\end{equation}
%
evaluating the characteristic spatial distance $\overline{x}(t)$ passed by the particle during the time interval $t$. Expression~\eqref{sec4:5} has been derived via the relationship between the probability functions 
\begin{equation*}
 \mathcal{P}(x,t) = \mathfrak{F}(\theta,\mathfrak{t}) \left[\frac{dx}{d\theta}\right]^{-1}
\end{equation*} 
and Exp.~\eqref{sec4:6} was obtained setting $\theta = \theta_\mathfrak{t}$ in Exp.~\eqref{sec4:4}.

The rigorous formula for the asymptotic behavior of the function $\mathcal{P}(x,t)$ was obtained in Ref.~\cite{we2} using a singular perturbation technique for $1<\alpha<2$ and then verified numerically also for $0<\alpha \leq 1$ \cite{we3}. Following Ref.~\cite{we3} we rewrite it as 
\begin{equation}\label{sec4:7}
   \mathcal{P}^\text{rig}(x,t) = \frac{4\alpha}{2^\alpha \Gamma^2(\alpha/2)}\cdot \frac{t}{x^{1+\alpha}}\,.
\end{equation}
Whence we see that the rigorous expression and the expression obtained using ansatz~\eqref{sec4:3} coincide with each other within the factor 
\begin{equation}\label{cofactor}
	\Omega(\alpha)= \left(\frac{2}{\alpha}\right)^\alpha \Gamma(1+\alpha)\in(1,2.05)
\end{equation}
for $\alpha\in(0,2)$. The fact that the obtained coefficient $\Omega(\alpha)$ is really about unity for all the values of the parameter $\alpha$ under consideration justifies Propositions~\ref{Prop5}--\ref{Prop9}.

\section{Conclusion and closing remarks}

The work has been devoted to the relationship between the continuous Markovian model for L\'evy flights developed previously \cite{we1,we2,we3,we33} and their equivalent representation in terms of discrete steps of a wandering particle. The present analysis has been confined to the one-dimensional model of continuous random motion of a particle with inertia. Its dynamics is studied in terms of random motion on the phase plane $\{x,v\}$ comprising the position $x$ and velocity $v=dx/dt$ of the given particle. Time variations in the particle velocity are considered to be governed by a stochastic differential equation whose regular term describes ``viscous'' friction with, maybe, a nonlinear friction coefficient $k(v)$.  Its stochastic term containing white Gaussian noise, the random Langevin force, allows for the stochastic self-acceleration phenomenon which can be of different nature. The stochastic self-acceleration is taken into account via the noise intensity $g(v)$ growing with the particle velocity $v$. Spacial attention is payed to the ideal case where the friction coefficient $k$ is constant and the noise intensity $g(v)\propto v$ becomes proportional to the particle velocity $v$ when the latter exceeds some threshold $v_a$. It is the case where the generated random walks exhibit the main properties of L\'evy flights. Namely, first, the distribution function $\mathcal{P}(x,t)$ of the particle displacement $x$ during the time interval $t$ possesses the power-law asymptotics. Second, the characteristic length $\overline{x}(t)$ of particle displacement during the time interval $t$ scales with $t$ also according to the power-law. 

The characteristic feature of the considered stochastic process is the fact that such nonlinear dependence of the noise intensity on the particle velocity gives rise to a multiscale time pattern $\{v(t)\}$ and the spatial particle displacement is mainly caused by the velocity extreme fluctuations, i.e., large peaks of the given pattern \cite{we1}. In particular, if we consider the velocity pattern  $\{v(t)\}$ of duration $t$ then the particle displacement $x$ within the corresponding time interval can be evaluated as $x\sim \vartheta_t \tau$, where $\vartheta_t$ is the amplitude of the largest peak available in the given pattern and $\tau$ is a ``microscopic'' time scale characterizing the velocity correlations. As a result, the statistical properties of the velocity fluctuations cause the L\'evy time scaling of the characteristic length $\overline{x}(t)$ of particle displacement and endow the distribution function $\mathcal{P}(x,t)$ with the appropriate power-law asymptotics. 

This feature has made it attractive to represent a trajectory of the wandering particle or, speaking more strictly, the time pattern $\{v(t)\}$ as a sequence of peaks of duration about $\tau$ and to consider each peak as a certain implementation of one discrete step of the particle motion. Unfortunately such an approach cannot be constructed directly because the time pattern  $\{v(t)\}$ as a random trajectory is not a smooth curve. To overcome this obstacle a complex neighborhood $\mathcal{L}$ of the line $v= 0$ on the phase plane $\{x,v\}$ has been introduced. It contains two boundaries $|v| = v_u$ and $|v| = v_l$, where the choice of the parameters $v_u$, $v_l$ meeting the inequality $v_l < u_u$ is determined by the simplicity of mathematical constructions. 
The presence of the two boundaries has enabled us to introduce the notion of random walks outside and inside the neighborhood $\mathcal{L}$.
The notion of random walks outside $\mathcal{L}$ describes a fragment of the particle motion in the region $|v|> v_l$ without touching the lower boundary $|v|=v_l$ until the particle gets it for the first time. Random walks inside the neighborhood $\mathcal{L}$ correspond to a fragment of the particle motion inside the region $|v|<v_u$ without touching the upper boundary $|v| = v_u$ again until the particle gets it for the first time. For the analyzed phenomena the initial particle velocity does not matter provided it is not too large. Therefore the initial particle velocity was set equal to $v_u$ without loss of generality. In this way any trajectory of the particle motion is represented as a sequence of alternate fragments of random walks inside and outside the neighborhood $\mathcal{L}$. A complex unit (basic unit) made of two succeeding fragments of the particle random motion inside and outside $\mathcal{L}$ may be treated as a continuous implementation of one step of the equivalent discrete random walks. The individual duration and the resulting length of these basic units are partly correlated random variables. It enables us to regard the constructed representation of random trajectories as a certain generalization of continuous time random walks (CTRW). The main difference between the CTRW model and the model developed here is the fact that the particle is not assumed to move uniformly along the straight line connecting the terminal points of one step.

For the analyzed model~\eqref{sec1:eq1} it has been demonstrated that the particle motion inside the neighborhood $\mathcal{L}$ practically contributes only to the duration of the basic units. The particle motion outside $\mathcal{L}$ determines the spatial displacement as well as contributes to the basic unit duration too. In the given model the kinetic coefficients, i.e., the friction coefficient $k(v)$ and the noise intensity $g(v)$ exhibit no singularities at $v=0$. Therefore the distribution of the basic unit duration has no anomalous properties and, thereby, there is a linear relationship
$$
    k = \gamma t
$$
between the running time $t$ and the number of basic units $k$, naturally, for $t\gg \tau$ and, so, for $k\gg 1$. The proportionality coefficient $\gamma$ has been obtained as a certain function of the model parameters.  As a result, in describing statistical properties of such random walks the system may be characterized by the number $k$ of basic units imposing no requirements on the duration of the pattern $\{v(t)\}$ as a whole. It should be pointed out that the developed classification of random trajectories holds within rather general assumptions about Markovian stochastic processes. So it can be generalized to models where the kinetic coefficients exhibit essential singularities in the region of small velocities. However, in this case there is no direct proportionality between $t$ and $k$, moreover, the integer $k$ has to be treated as a certain random variable; a similar situation is met in modeling grain boundary diffusion as a stochastic processes of the subdiffusion type \cite{Keigan1,Keigan2}. 

Using the constructed trajectory classification the analyzed stochastic process has been reduced to a certain universal stochastic process $\{\mathfrak{u(t)}\}$ for which the corresponding governing equation contains no parameters. Moreover, the white Gaussian noise enters this equation in the additive manner and the regular drift is a piece-wise constant function, namely, the $\sign(\mathfrak{u})$-function. It has been called the core stochastic process. 
All the basic parameters characterizing the generated random walks, e.g., the exponent of the L\'evy scaling law, are specified by the model parameters via the corresponding coefficients in the transformation from the variable $\mathfrak{u}$ of the core stochastic process $\{\mathfrak{u(t)}\}$ to the original particle velocity $v$. Namely, it is the transformation $\mathfrak{u}\to v$ as well as the linear transformation of the core process time to the ``physical'' time,  $\mathfrak{t}\to t$. This, in particular, explains us why the main results rigorously obtained for the superdiffusive regime of L\'evy flights hold also for the other possible regimes as demonstrated numerically \cite{we3}. 

In order to elucidate the basic properties of the random walks at hand the developed technique has been applied to the core stochastic process to construct the peak pattern $\mathbb{P}_\mathfrak{u}(k):=\{\mathfrak{u(t)}\}$ subjected to the condition $|\mathfrak{u(t)}|<\theta$ for all the moments of time, where $\theta\gg1$ is a certain given value. Based on the probabilistic properties of the core stochastic process it has enabled us also to introduce the probability density $\mathfrak{F(\theta,t)}$ that the maximum attained by the variable $|\mathfrak{u(t)}|$ inside the pattern $\mathbb{P}_\mathfrak{u}(k)$ is equal to $\theta$. The probability density $\mathfrak{F(\theta,t)}$ has been studied in detail in the present paper. In particular, first, it has been found out that the region of the asymptotic behavior of $\mathfrak{F(\theta,t)}$ is specified by the inequality $\theta \gtrsim \theta_\mathfrak{t}$, where $\theta_\mathfrak{t} = \ln(\mathfrak{t})$. Second, the so-called  single-peak approximation has been justified. It states that if the value $\theta \gtrsim \theta_\mathfrak{t}$ is attained within a given peak of the pattern $\mathbb{P}_\mathfrak{u}(k)$ then the variations of the variable $\mathfrak{u}$ inside the other peaks may be assumed to be small in comparison with $\theta$ and the condition $|\mathfrak{u}|\leq \theta$ does not really affect their statistics.
Then, as been demonstrated, the asymptotic properties of the generated random walks can be formulated in terms of the core stochastic process as follows. 

\begin{itemize}
\item On time scales $t\gg\tau$ large fluctuations in the spatial displacement $x$ of the wandering particle are implemented mainly via its motion within the single peak of the pattern $\mathbb{P}_\mathfrak{u}(k)$ whose  amplitude is maximal in comparison with the other peaks and attains extremely large values.

\item  The region $|x|\gg \overline{x}(t)$ of large fluctuations in the particle spatial displacement $x$ matches directly the region $\theta\gtrsim\theta_\mathfrak{t}$ of the extreme values $\theta$ attained by the variable $\mathfrak{u(t)}$ inside the pattern $\mathbb{P}_\mathfrak{u}(k)$.

\item The asymptotic behavior of the distribution function $\mathcal{P}(x,t)$ in the region $|x|\gg \overline{x}(t)$ is practically specified by the asymptotic behavior of the probability density $\mathfrak{F}(\theta,\mathfrak{t})$ inside the region  $\theta\gtrsim\theta_\mathfrak{t}$.

\item The lower boundary $\overline{x}(t)$ of the asymptotic behavior of $\mathcal{P}(x,t)$ determines the characteristic displacement of the wandering particle during the time interval $t$. So it can be evaluated using the single-peak approximation and setting the value $\theta$ equal $\theta_\mathfrak{t}$, i.e., $\theta\sim \theta_\mathfrak{t}$. It should be reminded that $\theta_\mathfrak{t}$ is also the lower boundary of the $\theta$-region wherein the single-peak approximation holds.
\end{itemize}

In addition it is worthwhile to note that, first, in studying the extreme characteristics of particle motion outside the neighborhood $\mathcal{L}$ the shape of the velocity pattern $\{v(t)\}$ can be approximated using the most probable trajectory $\{\mathfrak{u_\text{opt}(t)}\}$. In particular using these results the asymptotics of the distribution function $\mathcal{P}(x,t)$ has be constructed and demonstrated to coincide with the rigorous results within a cofactor about unity. It opens a gate to modeling such processes in heterogeneous media constructing the most optimal trajectories of particle motion in nonuniform environment. 

Second, the first item above can be regarded as a certain implementation of the single-peak approximation. It is based on the use of the condition $|\mathfrak{u(t)}|<\theta$ in constructing the pattern  $\mathbb{P}_\mathfrak{u}(k)$. However, the developed classification of random trajectories admits a more sophisticated analysis of anomalous stochastic processes. To do this it is necessary to consider a more complex system of restrictions $|\mathfrak{u(t)}|_i<\theta_i$ with the independent external boundaries $\{\theta_i\}$ for different peaks. In this case it would be possible to speak about several peaks of different large amplitude and to go beyond the single-peak approximation.

\section*{Acknowledgments}

The work was supported in part by the JSPS ``Grants-in-Aid for Scientific Research'' Program, Grant 245404100001, as well as the Competitive Research Funding of the University of Aizu, Project P-25, FY2012. 

\appendix

\section{Probabilistic properties of random walks inside and outside the layer $\mathcal{L}$}\label{App:inL}

It should be noted beforehand that in the present Appendix no approximation of the potential $\Phi(\eta)$ introduced by expression~\eqref{sec1:Phi} will be used, only the general properties \eqref{sec1:kgaa} of the kinetic coefficients $k(v)$ and $g(v)$ are taken into account. It enables us to make use of the results to be obtained here in further generalizations, e.g., to allow for the cutoff effects.

\subsection{Terminal fragments: the limit case $s\to 0$ and $\theta\to\infty$}\label{AppG}

The limit $\theta\to\infty$ describes the situation when the upper boundary $\eta=\theta$ of the analyzed region $[0,\theta)\ni\eta$ is placed rather far away from the origin $\eta=0$ and its effect on the random particle motion is ignorable. In this case, from the general point of view, the terminal fragments shown in Fig.~\ref{F2} are no more than random walks starting from at a given point $\eta_0$ and reaching another point $\eta$ in a  time $t$ without touching a certain boundary $\eta=\zeta$. Their probabilistic properties are described, in particular, by the probability density $\mathcal{G}(\eta,t|\eta_0,\zeta)$ of finding the random walker at the point $\eta$ in the time $t$. The Laplace transform $G(\eta,s|\eta_0,\zeta)$ of this function obeys the following forward Fokker-Planck equation matching the Langevin equation~\eqref{sec1:eq2} (see, e.g., Ref.~\cite{Gardiner})
\begin{equation}\label{AppG:fFP}
 s G = \frac{\partial}{\partial \eta}\left[\frac{\partial G}{\partial \eta} + \alpha\frac{d\Phi(\eta)}{d\eta} G\right] + \delta(\eta-\eta_0)\,.
\end{equation} 
For random walks inside the layer $\mathcal{L}_\zeta=[0,\zeta)$ with the initial point $\eta_0 < \zeta$  Eq.~\eqref{AppG:fFP} should be subjected to the boundary conditions 
\begin{subequations}\label{AppG:0U}
\begin{align}\label{AppG:0Ua}
   \left[\frac{\partial G}{\partial \eta}+\alpha\frac{d\Phi}{d\eta} G\right]_{\eta = 0} & = 0\,, &
    \left. G \right|_{\eta = \zeta} & = 0\,.
\\
\intertext{For random walks outside the layer $\mathcal{L}_\zeta$ with  $\eta_0>\zeta$ the corresponding boundary conditions are} 
\label{AppG:0Ub}
    \left[\frac{\partial G}{\partial \eta}+\alpha\frac{d\Phi}{d\eta}G\right]_{\eta \to \infty} & \to 0\,, &
    \left. G \right|_{\eta = \zeta}& = 0\,.
\end{align}
\end{subequations}
Since the effect of time in the analyzed phenomena is mainly caused by the properties of the peak pattern $\mathbb{P}(t|k)$ and the time scales $t\gg1$ are of the primary interest, dealing with the terminal fragments we may confine our consideration to the limit $s\to0$. In this case Eq.~\eqref{AppG:fFP} is reduced to the following   
\begin{equation}\label{AppG:fFP0}
 \frac{\partial}{\partial \eta}\left[\frac{\partial G}{\partial \eta} + \alpha\frac{d\Phi(\eta)}{d\eta} G\right] = - \delta(\eta-\eta_0)
\end{equation} 
and after simple mathematical manipulations using the method of variation of constants for solving difference equations we get the desired expression  for random walks inside the layer $\mathcal{L}_\zeta$
\begin{subequations}\label{AppG:GL}
\begin{align}
\label{AppG:GLin}  
    G^\text{in}(\eta|\eta_0,\zeta):= G(\eta,s|\eta_0,\zeta)\big|_{\substack{\eta_0<\zeta\\ s\to0}}
                                         & = e^{-\alpha\Phi(\eta)}\int\limits_{\eta_0}^\zeta e^{\alpha\Phi(\eta')} \Theta_\text{H}(\eta'-\eta)\,d\eta'
\\ 
\intertext{and for random walks outside the layer $\mathcal{L}_\zeta$}
\label{AppG:GLout}
    G^\text{out}(\eta|\eta_0,\zeta) := G(\eta,s|\eta_0,\zeta)\big|_{\substack{\eta_0>\zeta\\ s\to0}} 
                                         & = e^{-\alpha\Phi(\eta)}\int\limits^{\eta_0}_\zeta e^{\alpha\Phi(\eta')} \Theta_\text{H}(\eta-\eta')\,d\eta'\,,
\end{align}
\end{subequations}
where $\Theta_\text{H}(\ldots)$ is the Heaviside step function determined by Exp.~\eqref{sec2:Heav}. 

The first terminal fragment $\mathbb{G}^\text{in}_n(\eta)$ (Fig.~\ref{F2}) matches the analyzed random walks starting at the point $\eta_l$ and reaching the point $\eta$ in the time $\Delta t = t-t_n$ without touching the boundary $\eta=\eta_u > \eta_l$. So the probabilistic weight $\mathcal{G}^\text{in}(\eta,\Delta t)$ of this fragment is specified by the expression
\begin{equation*}
    \mathcal{G}^\text{in}(\eta,\Delta t) = \mathcal{G}(\eta,\Delta t|\eta_0,\zeta)\big|_{\substack{\eta_0=\eta_l\\\zeta=\eta_u}}
\end{equation*}
and its Laplace transform ${G}^\text{in}(\eta,s)$ in the limit $s\to0$ takes the form
\begin{subequations}\label{AppG:GLFinal}
\begin{equation}\label{AppG:GLinFinal}  
    G^\text{in}(\eta,s)\big|_{s\to0} = e^{-\alpha\Phi(\eta)}\int\limits_{\eta_l}^{\eta_u} e^{\alpha\Phi(\eta')} \Theta(\eta'-\eta)\,d\eta'
\end{equation}
by virtue of \eqref{AppG:GLin}.
The second terminal fragment $\mathbb{G}^\text{out}_n(\eta)$ (Fig.~\ref{F2}) is also represented by the given random walks starting at the point $\eta_u$ and reaching the point $\eta$ in the time $\Delta t = t-t_n$ without touching the boundary $\eta_l$. Thereby the Laplace transform ${G}^\text{out}(\eta,s)$ of its probabilistic weight $\mathcal{G}^\text{out}(\eta,\Delta t)$ is specified by Exp.~\eqref{AppG:GLout}, namely,
\begin{equation}\label{AppG:GLoutFinal}  
    G^\text{out}(\eta,s)\big|_{s\to0} = e^{-\alpha\Phi(\eta)}\int\limits_{\eta_l}^{\eta_u} e^{\alpha\Phi(\eta')} \Theta(\eta-\eta')\,d\eta'\,.
\end{equation}
\end{subequations}
It should be noted that in both of Exps.~\eqref{AppG:GLFinal} the quantity $\eta$ can take any arbitrary value $\eta\in (0,\infty)$ rather than a value from the corresponding interval only. Indeed, if the taken value falls outside this interval the relevant expression will give out the probability density equal to zero.

The latter feature enables us to represent the construction reproducing actually the right-hand side of Exp.~\eqref{sec2:G0} as 
\begin{subequations}\label{AppG:GGFinal}
\begin{multline}\label{AppG:GGFinal1}
   \Theta_\text{H}(\eta -\eta_l) G^\text{out}(\eta,s)\big|_{s\to0} + \Theta_\text{H}(\eta_u -\eta) G^\text{in}(\eta,s)\big|_{s\to0} =
\\
    G^\text{out}(\eta,s)\big|_{s\to0} + G^\text{in}(\eta,s)\big|_{s\to0} = e^{-\alpha\Phi(\eta)}\int\limits_{\eta_l}^{\eta_u} e^{\alpha\Phi(\eta')} d\eta'\,.
\end{multline}
In deriving this expression the identity $\Theta(\eta-\eta')+ \Theta(\eta'-\eta) \equiv 1$ has been taken  into account. In the cause under consideration it is assumed that the potential $\Phi(\eta)$ can be approximated by a linear function of $\eta$ in the region $\eta\sim 1$, namely, $\Phi(\eta)\approx C + \eta$, where $C$ is some constant. Under such conditions the previous expression is reduced to  
\begin{equation}\label{AppG:GGFinal2}
   \Theta_\text{H}(\eta -\eta_l) G^\text{out}(\eta,s)\big|_{s\to0} + \Theta_\text{H}(\eta_u -\eta) G^\text{in}(\eta,s)\big|_{s\to0} = \frac1{\alpha}
    e^{-\alpha\Phi(\eta)}\left[e^{\alpha\Phi(\eta_u)} - e^{\alpha\Phi(\eta_l)} \right]\,.
\end{equation}
\end{subequations}
In particular, these expressions demonstrate that in the limit $s\to 0$, i.e., when $t\to\infty$ the cumulative contribution of the two terminal fragments to the probability density of finding the random walker at the point $\eta$  is equal to the steady state distribution of the random variable $\eta$ within a certain factor of proportionality.

\subsection{Fragments of the peak pattern $\mathbb{P}(t\mid k)$}\label{AppF}

Generally the individual fragments of the peak pattern represented  in Fig.~\ref{F3} can be regarded as random walks of a particle that initially ($t=0$) is located at a point $\eta$ and gets a boundary $\zeta$ for the first time at a moment $t$. If the initial point $\eta$ is located outside the interval $\mathcal{L}_\zeta := [0,\zeta)$, i.e., $\eta >\zeta$, then the addition condition is imposed on the particle motion; it is not allowed for it to touch or cross the distant boundary $\theta\gg1$. The probabilistic properties of these fragments are described by the probability density $\mathcal{F}(\zeta,t|\eta,\theta)$ that the particle starting from the point $\eta$ at $t=0$ gets the boundary $\zeta$ at the moment $t$ for the first time and, in addition when applicable, during its motion never crosses the distant boundary $\theta$.   

The Laplace transform ${F}(\zeta,s|\eta,\theta)$ of this function determined by Exp.~\eqref{sec2:Lapl} obeys the following backward Fokker-Planck equation matching the Langevin equation~\eqref{sec1:eq2} (see, e.g., Ref.~\cite{Gardiner})  
\begin{equation}\label{AppF:bFP}
  s F = \frac{\partial^2 F}{\partial \eta^2} - \alpha\frac{d\Phi(\eta)}{d\eta} \frac{\partial F}{\partial \eta}\,.
\end{equation} 
For the random walks inside the layer $\mathcal{L}_\zeta$, i.e. when $0\leq\eta<\zeta$ Eq.~\eqref{AppF:bFP} should be subjected to the boundary conditions  
\begin{subequations}\label{AppB:BC}
\begin{align}
\label{AppF:BCin}
  F\big|_{\eta = \zeta} & = 1\,, & \left.\frac{\partial F}{\partial \eta}\right|_{\eta = 0} & = 0\,,\\
\intertext{and for the random walks outside the layer $\mathcal{L}_\zeta$, i.e. when $\zeta<\eta<\theta$ the relevant boundary conditions are}
\label{AppF:BCOut} 
  F\big|_{\eta = \zeta} & = 1\,, & F\big|_{\eta = \theta}& = 0\,.
\end{align}
\end{subequations}
Since the details of solving Eq.~\eqref{AppF:bFP} for the random walks inside and outside the layer $\mathcal{L}_\zeta$ are different we will analyze the two cases individually assuming the inequality $s\ll1$ to hold beforehand, which matches large time scales $t\gg1$.
   
\subsubsection{Random walks inside the layer $\mathcal{L}_\zeta$}\label{AppFIN}

Because values of $\eta_l$ and $\eta_u$ about unity are of the prime interest, see the corresponding discussion in Sec.~\ref{sec:TC}, here we may consider the thickness of the layer $\mathcal{L}_\zeta$ to be also about unity, $\zeta\sim 1$. In this case at the zero-th approximation in $s$ the solution of Eq.~\eqref{AppF:bFP} subject to the boundary conditions~\eqref{AppF:BCin} is equal to unity. So at the first approximation in $s$ Eq.~\eqref{AppF:bFP} can be rewritten as
\begin{equation}\label{AppF:bFPin}
  s  = \frac{\partial^2 F}{\partial \eta^2} - \alpha\frac{d\Phi(\eta)}{d\eta} \frac{\partial F}{\partial \eta}\,.
\end{equation} 
Using the method of variation of constants Eq.~\eqref{AppF:bFPin} under conditions~\eqref{AppF:BCin} is integrated directly, yielding the desired expression
\begin{equation}\label{AppF:FGin}
 F^\text{in}(\zeta,s|\eta) = 1- s \int\limits_\eta^{\zeta}d\eta'\, e^{\alpha\Phi(\eta')} \int\limits_0^{\eta'}d\eta'' e^{-\alpha\Phi(\eta'')}\,.
\end{equation}
Here the parameter $\theta$ has been omitted from the list of arguments and the superscript `in' has been added to denote the analyzed region explicitly. 

Expression~\eqref{AppF:FGin} immediately enables us to write the Laplace transform of the individual probabilistic weights $\mathcal{F}^\text{in}(\Delta t)$ of the fragments  $\mathbb{F}^\text{in}_{i,i+1}:= \mathbb{F}^\text{in}(\Delta t)$ with $\Delta t = t_{i+1}-t_i$ (Fig.~\ref{F2}) in the form
\begin{equation}\label{AppF:Fin}
 F^\text{in}(s) = 1- s \int\limits_{\eta_l}^{\eta_u}d\eta'\, e^{\alpha\Phi(\eta')} \int\limits_0^{\eta'}d\eta'' e^{-\alpha\Phi(\eta'')}
\end{equation}
within the approximation of large time scales $t\gg1$ or, equivalently, for $s\ll1$. 

\subsubsection{Random walks outside the layer $\mathcal{L}_\zeta$}

Since the analyzed phenomena are governed by large fluctuations in the particle velocity, the upper boundary of the region under consideration $\eta\in(\zeta,\theta)$ is presumed to be a rather distant point, $\theta\gg1$, in addition to the assumption $s\ll1$. Then let us seek the desired solution of Eq.~\eqref{AppF:bFP} subject to the boundary conditions~\eqref{AppF:BCOut} in the form
\begin{equation}\label{AppF:FF}
   F(\eta) = A_s F_s(\eta)+ A_0 F_0(\eta)\,,
\end{equation}
where $A_s$, $A_0$ are some constants and the functions $F_s(\eta)$, $F_0(\eta)$ are specified via the expression
\begin{equation}\label{AppF:Fk}
    F_{s,0}(\eta) = \exp\left\{
    \int\limits_{\zeta}^\eta k_{s,0}(\eta')\,d\eta'
    \right\}\,.
\end{equation}
The boundary conditions~\eqref{AppF:BCOut} imposed on solution~\eqref{AppF:FF} enable us to calculate the coefficients $A_s$, $A_0$, and, then, to rewrite solution~\eqref{AppF:FF} in the form
\begin{equation}\label{AppF:FFfin}
   F(\eta) = \frac{F_0(\theta)F_s(\eta) - F_s(\theta)F_0(\eta)}{F_0(\theta) - F_s(\theta)}  
           = F_s(\eta)- \frac{F_s(\theta)}{F_0(\theta)-F_s(\theta)}\cdot\left[F_0(\eta)-F_s(\eta)\right]\,.
\end{equation}
In deriving Exps.~\eqref{AppF:FFfin} the identity $F_0(\zeta)=F_s(\zeta)=1$ stemming directly from definition~\eqref{AppF:Fk} has been taken into account.

The substitution of \eqref{AppF:Fk} into \eqref{AppF:bFP} shows the function $k(\eta):=k_{s,0}(\eta)$ to obey the Riccati equation
\begin{equation}\label{AppF:Ric}
    s = k^2 +\frac{dk}{d\eta} -\alpha\frac{d\Phi(\eta)}{d\eta} k\,.
\end{equation}
In the region $\eta\gtrsim 1$ (when also $\zeta\gtrsim1$) the potential $\Phi(\eta)$ introduced via Exp.~\eqref{sec1:Phi} is approximately a linear function of its argument, namely, $\Phi(\eta)\approx\eta + C$, where $C$ is a certain constant, by virtue of the accepted assumption~\eqref{sec1:kgaa}. In this case $d\Phi(\eta)/d\eta \approx 1$ and the Riccati equation~\eqref{AppF:Ric} has two solutions 
\begin{align}
\label{AppF:ks:0}
    k_s(\eta)& \approx -\frac{s}{\alpha}\,,
\\
\label{AppF:k0:0} 
    k_0(\eta)& \approx \alpha\,,
\end{align}
written in the approximation of leading terms in $s\ll1$.
Since we can choose any two \textit{specific independent} solutions of the Riccati equation~\eqref{AppF:Ric} let us impose on the desired solutions  $k=k_{s,0}(\eta)$ the requirement that in the region $\eta\gtrsim1$ both of them tend to the constant values~\eqref{AppF:ks:0}, \eqref{AppF:k0:0}, respectively.

In accordance with the results to be obtained the solution $k_s(\eta)$ can be treated as a small quantity of order $s$ due to the monotonous growth of the potential $\Phi(\eta)$ with its argument $\eta$. It enables us to omit the quadratic term from the Riccati equation~\eqref{AppF:Ric} and, then, using the method of variation of constants write the desired solution as
\begin{equation}\label{AppF:ks:1}
    k_s(\eta) = -s e^{\alpha\Phi(\eta)}\int\limits_\eta^\infty e^{-\alpha\Phi(\eta')}\,d\eta'\,.
\end{equation}
Whence it follows, in particular, that the function $F_s(\eta)$ is decreasing one and its asymptotic behavior as $s\to 0$ is specified by the expression
\begin{equation}\label{AppF:FonS}
  F_s(\eta) = 1 - s\int\limits_\zeta^\eta d\eta' e^{\alpha\Phi(\eta')}\int\limits_{\eta'}^\infty d\eta'' e^{-\alpha\Phi(\eta'')}\,.
\end{equation}
For the analyzed phenomena we need to know only the characteristics of small deviations of the function $F_s(\eta)$ from unity for $s\ll 1$, which actually has enabled us to construct the function $k_s(\eta)$ within the accuracy of the leading term in $s$.

In constructing the second solution $k_0(\eta)$ we may set the left-hand side of equation~\eqref{AppF:Ric} equal to zero and, then, rewrite it as follows
\begin{equation}\label{AppF:k0:1}
    k = \alpha\frac{d\Phi(\eta)}{d\eta} -\frac{d\ln(k)}{d\eta} \,.
\end{equation}
In the case under consideration the deviation of the function $\Phi(\eta)$ from the linear dependence can occur only on scales $\eta\gg1$. So we may confine ourselves to the first iteration in \eqref{AppF:k0:1}, yielding us the desired approximation 
\begin{equation}\label{AppF:k0:2}
      k_0(\eta) = \alpha\frac{d\Phi(\eta)}{d\eta} -\frac{d}{d\eta}\ln\left[\frac{d\Phi(\eta)}{d\eta}\right]\,.
\end{equation}
Expression~\eqref{AppF:k0:2} and definition~\eqref{AppF:Fk} immediately enable us to specify the function $F_0(\eta)$, namely,
\begin{equation}\label{AppF:F0}
  F_0(\eta) = \left[\frac{d\Phi(\eta)}{d\eta}\right]^{-1} e^{\alpha\left[\Phi(\eta)-\Phi(\zeta)\right]}\,.
\end{equation}
In obtaining this formula we have set $d\Phi(\eta)/d\eta=1$ at $\eta=\zeta\gtrsim 1$. In particular, this formula demonstrates us that the function $F_0(\eta)$ exhibits the exponential growth as the variable $\theta$ runs to large values. So the contribution of the finite magnitude of the parameter $s\ll 1$ to the function $F_0(\eta)$ becomes essential only when the argument $\eta$ attains such large values that the function $F_0(s)$ even in the leading approximation in $s$, i.e., for $s=0$ has already got exponentially large values. This region is of minor interest in the frameworks of the present analysis, which has enabled us to construct the solution $k_0(\eta)$ setting $s=0$. 

For the same reasons we may confine ourselves to approximation~\eqref{AppF:FonS} in evaluating the first term on the right-hand side of Exp.~\eqref{AppF:FFfin} and in the second term set $F_s(\eta) = F_s(\theta) = 1$. It is due to the fact that for the most interesting magnitudes of the parameter $\theta$ the effects caused by the finite magnitude of the parameter $s$ are ignorable.  In this way using formulas~\eqref{AppF:FonS} and \eqref{AppF:F0} we gets the approximation
\begin{equation}\label{AppF:Fst}
    F(\eta) = 1 - s \int\limits_\zeta^\eta d\eta' e^{\alpha\Phi(\eta')}\int\limits_{\eta'}^\infty d\eta''\, e^{-\alpha\Phi(\eta'')}
                     - \left.\frac{d\Phi(\eta)}{d\eta}\right|_{\eta=\theta}\left[\frac{e^{\alpha\Phi(\eta)}-e^{\alpha\Phi(\zeta)}}{e^{\alpha\Phi(\theta)}}\right]\,.
\end{equation}
Setting here $\eta=\eta_u$ and $\zeta = \eta_l$ we immediately get the desired expression for the Laplace transform  of the individual probabilistic weights $\mathcal{F}^\text{out}(\Delta t)$ of the fragments $\mathbb{F}^\text{out}_{i,i+1}:= \mathbb{F}^\text{out}(\Delta t)$ with $\Delta t = t_{i+1}- t_i$ (Fig.~\ref{F2}) in the form
\begin{equation}\label{AppF:Fout}
      F^\text{out}(s) = 1 - s \int\limits_{\eta_l}^{\eta_u} d\eta' e^{\alpha\Phi(\eta')}\int\limits_{\eta'}^\infty d\eta''\, e^{-\alpha\Phi(\eta'')}
                     - \left.\frac{d\Phi(\eta)}{d\eta}\right|_{\eta=\theta}\left[\frac{e^{\alpha\Phi(\eta_u)}-e^{\alpha\Phi(\eta_l)}}{e^{\alpha\Phi(\theta)}}\right]\,.
\end{equation}
It should be reminded that Exp.~\eqref{AppF:Fout} as well as \eqref{AppF:Fst} assumes the estimate $d\Phi(\eta)/d\eta=1$ to hold for $\eta,\eta_l,\eta_u,\zeta\gtrsim 1$.

\subsubsection{Composed element of the peak pattern $\mathbb{P}(t\mid k)$}

The construction of the peak pattern $\mathbb{P}(t\mid k)$, see Fig.~\ref{F3} and Exp.~\eqref{sec2:G2}, is based on the repetition of the same unit, a composed element made of random walks outside the region $\mathcal{L}$ followed by random walks inside it. The probabilistic weight $\mathcal{F}(t)$ of this unit is determined by its Laplace transform equal to the product
\begin{equation}\label{AppF:Unit1}
   F(s) = F^\text{in}(s)\cdot F^\text{out}(s)\,.
\end{equation}
Whence taking into account formulas~\eqref{AppF:Fin} and \eqref{AppF:Fout} we immediately get the expression 
\begin{subequations}\label{AppF:Unit2}
\begin{multline}\label{AppF:Unit2a}
      F(s) = 1 - s \int\limits_{\eta_l}^{\eta_u} d\eta' e^{\alpha\Phi(\eta')}\int\limits_{0}^\infty d\eta''\, e^{-\alpha\Phi(\eta'')}
                     - \left.\frac{d\Phi(\eta)}{d\eta}\right|_{\eta=\theta}\left[\frac{e^{\alpha\Phi(\eta_u)}-e^{\alpha\Phi(\eta_l)}}{e^{\alpha\Phi(\theta)}}\right]
\\
    {}=1  - \frac1{\alpha}\left[{e^{\alpha\Phi(\eta_u)}-e^{\alpha\Phi(\eta_l)}}\right] \left[{s} \int\limits_{0}^\infty d\eta''\, e^{-\alpha\Phi(\eta'')}
                         + {\alpha}\left.\frac{d\Phi(\eta)}{d\eta}\right|_{\eta=\theta}{e^{-\alpha\Phi(\theta)}}\right]\,.
\end{multline}
Using formula~\eqref{petast} for the stationary distribution function $p^\text{st}(\eta)$ of the random variable $\eta$ and actually merging the half-spaces $\{\eta>0\}$ and $\{\eta <0\}$ the obtained expression is reduced to the following
\begin{equation}\label{AppF:Unit2b}
      F(s) =1  - \frac1{\alpha}\left[{e^{\alpha\Phi(\eta_u)}-e^{\alpha\Phi(\eta_l)}}\right] 
      \left[{s} - \left.\frac{dp^\text{st}(\eta)}{d\eta}\right|_{\eta=\theta}\right]\int\limits_{0}^\infty d\eta'\, e^{-\alpha\Phi(\eta')}\,.
\end{equation}
\end{subequations}
It gives us the desired probabilistic weight in the frameworks of the analyzed case, i.e., for large time scales, $t\gg 1$, or equivalently $s\ll 1$, the distant upper boundary of the region of random walks, $\theta \gg 1$, and the accepted assumption about the behavior of the potential $\Phi(\eta)$ in the region $\eta\sim 1$, namely, $\Phi(\eta) \approx C + \eta$, where $C$ is a constant.

\end{document}